\documentclass[a4paper,10pt]{article}

\usepackage{amsmath}
\usepackage{amssymb}
\usepackage{amsfonts}
\usepackage{amscd}
\usepackage{latexsym}
\usepackage{amsbsy}
\usepackage{bbm}
\usepackage[english]{babel}
\usepackage{graphicx}
\usepackage{psfrag}

\begin{document}

 \begin{flushright}
 \small
 UG-09-25\\
 \today
 \normalsize
 \end{flushright}

 \begin{center}

 \vspace{.5cm}

 {\Large {\bf BPS Open Strings and $A$-$D$-$E$-singularities}}\\[.5cm]
 {\Large {\bf in F-theory on K3}}

 \vspace{.7cm}

 \begin{center}

 {\bf Eric A.~Bergshoeff ${}^\dagger$ and  Jelle Hartong} ${}^\star$\\[0.2cm]

 \vskip 25pt

 ${}^\dagger$ {\em Centre for Theoretical Physics, University of Groningen, \\
 Nijenborgh 4, 9747 AG Groningen, The Netherlands \vskip 5pt }

 {email: {\tt E.A.Bergshoeff@rug.nl }} \\
 \vskip 15pt

 ${}^\star$ {\em  Albert Einstein Center for Fundamental Physics,\\
 Institute for Theoretical Physics,\\
 University of Bern,\\
 Sidlerstrasse 5, 3012 Bern,
 Switzerland\vskip 5pt}

 {email: {\tt hartong@itp.unibe.ch}}
 \vskip 15pt

 \end{center}

 \vspace{1cm}

 {\bf Abstract}

 \begin{quotation}

 {\small
We improve on a recently constructed graphical representation of the supergravity 7-brane solution and apply this refined representation to re-study the open string description of the  $A$-$D$-$E$-singularities in F-theory on K3.
A noteworthy feature of the graphical representation is that it provides the complete
global branch cut structure of the 7-brane solution which plays an important role in our analysis.
We first identify those groups of branes which when made to coincide lead to the $A$-$D$-$E$-gauge groups. We next show that there is always a sufficient number of open BPS strings to account for all the generators of the gauge group. However, as we will show, there is in general no one-to-one relation between BPS strings and gauge group generators.

For the $D_{n+4}$- and $E$-singularities, in order to relate BPS strings with gauge group generators, we make an $SU(n+4)$, respectively $SU(5)$ subgroup of the $D_{n+4}$- and $E$-gauge groups manifest. We find that only for the $D$-series (and for the standard $A$-series) this is sufficient to identify, in a one-to-one manner, which BPS strings correspond to which gauge group generators.

}

 \end{quotation}

 \end{center}

 \newpage

 \section{Introduction}

Recently F-theory compactifications on a Calabi--Yau 4-fold down to four dimensions have been argued to have interesting phenomenological features \cite{Beasley:2008dc,Beasley:2008kw}. In these compactifications the Calabi-Yau 4-fold is described by an elliptic fibration of a K3 over a complex 2-dimensional surface $S$. This surface $S$ is wrapped by 7-branes which are space-time filling in the non-compact dimensions. The gauge groups of the wrapped 7-branes can be obtained geometrically through the $A$-$D$-$E$-singularities of the elliptically fibered K3. Special attention has
been given to the case of exceptional gauge groups.

Seven-branes with exceptional gauge groups are poorly understood objects in type IIB string theory. We will study the symmetry enhancement from an open string point of view making use of a recent analysis of the simplest (flat world-volume) supergravity 7-brane solutions \cite{Bergshoeff:2006jj}, which emphasizes the supersymmetry properties of the solutions. This analysis has yielded a graphical representation of the 7-brane solutions that summarizes the global branch cut structure of the two analytic functions in terms of which the entire solution can be described. Here we will improve on these graphical representations in a way that allows us to study open strings in a background of 7-branes. The analysis will be concerned with the case of 24 (flat world-volume) 7-branes forming F-theory on K3 \cite{Vafa:1996xn}.

Open string descriptions of the $A$-$D$-$E$-singularities have already been studied a long time ago \cite{Johansen:1996am,Gaberdiel:1997ud,Gaberdiel:1998mv}.
The motivation to re-study the open string description of these singularities is that with the work of \cite{Bergshoeff:2006jj} we now have
a full knowledge of the global branch cut structure of the complex axi-dilaton field $\tau$. This has led us to a different approach of the
problem in which we avoid the use of the so-called B- and C-branes \cite{Johansen:1996am,Gaberdiel:1997ud,Gaberdiel:1998mv}. In our picture these branes are represented by (1,0) 7-branes which are hidden behind $S$-branch cuts. These $S$ branch cuts play an important role in our analysis.

\section{Seven-branes: a short review}\label{sec:sevenbranes}

We will start in Subsection \ref{subsec:solutions} with a short review of the 24 7-brane solution. The $A$-$D$-$E$-singularities are next discussed in Subsection \ref{subsec:ADEsingularities}. In Subsection \ref{subsec:graphical} we discuss the graphical representation \cite{Bergshoeff:2006jj} of the 7-brane solution presented in Subsection \ref{subsec:solutions}. The notion of 7-brane charges in the global solution is discussed in Subsection \ref{subsection:charges}. This section ends with Subsection \ref{subsec:othersolutions} where we make some comments regarding other 7-brane solutions.

\subsection{Solutions}\label{subsec:solutions}

The basic 7-brane solution with a compact transverse space  requires 24 non-coincident 7-branes \cite{Greene:1989ya,Gibbons:1995vg}. This transverse space has the topology of $S^2$ with 24 punctures, the locations of the 24 7-branes. The solution is described by two analytic functions $\tau(z)$ and $f(z)$ in terms of which the metric (in Einstein frame) and the Killing spinor are given by
\begin{eqnarray}
 ds^2 & = & -dx_{1,7}^2+\text{Im}\,\tau\vert f\vert^2dzd\bar z\,,\label{IIBbackgroundmetric}\\
\epsilon &=&\left(\frac{\bar f}{f}\right)^{1/4}\epsilon_0\,,\hskip 1truecm \text{with}
\hskip 1truecm \gamma_z\epsilon_0=0\,,\label{Killingspinor}
\end{eqnarray}
for some constant spinor $\epsilon_0$. The metric $dx_{1,7}^2$ denotes
8-dimensional Minkowski space-time. The 7-brane transverse space is
parametrized in terms of $z,\bar z$ which are fixed up to an
$SL(2,\mathbb{C})$ coordinate transformation. The solution preserves 16 supersymmetries provided that the Killing
spinor $\epsilon$ is given by Eq. \eqref{Killingspinor}
\cite{Bergshoeff:2006jj}\footnote{The conventions for the unbroken
supersymmetries we use here are slightly different from the ones
used in  \cite{Bergshoeff:2006jj}.}. The holomorphic functions $\tau(z)$ and $f(z)$ are given by
\begin{eqnarray}
j(\tau) & = & \frac{P_8^3}{P_8^3+Q_{12}^2}\,,\label{IIBbackgroundtau}\\
f(z) & = & c\,\eta^2(\tau)\left(P_8^3+Q_{12}^2\right)^{-1/12}\,,\label{IIBbackgroundf}
\end{eqnarray}
for some nonzero complex constant $c$. The functions $j$ and
$\eta$ are Klein's modular $j$-function and the Dedekind
eta-function, respectively. Furthermore, $P_8$ and $Q_{12}$ are
arbitrary polynomials of degree $8$ and $12$, respectively, in the
complex coordinate $z$.

The complex axi-dilaton field can be interpreted as the modulus of a
2-torus that is elliptically fibered over the 7-brane transverse
space. If we describe this torus locally via a complex coordinate
$w$ then the complex 2-dimensional space parameterized in terms of
$z$ and $w$ forms a K3 surface \cite{Greene:1989ya,Vafa:1996xn}. The function $f$ has the
interpretation of $fdzdw$ being the holomorphic (2,0) form of the K3
\cite{Greene:1989ya}.

We have the following scale transformation (with complex parameter $\lambda$):
\begin{equation}
 P_8\rightarrow \lambda^2P_8\,,\qquad Q_{12}\rightarrow\lambda^3Q_{12}\,.
\end{equation}
This transformation leaves $j(\tau)$ invariant and provided we replace $c\rightarrow\lambda^{1/2}c$ it also leaves $f$ invariant. If we combine this scale transformation with the $SL(2,\mathbb{C})$ coordinate freedom it is concluded that we can fix at will four complex parameters that appear in the polynomials $P_8$ and $Q_{12}$. Since $P_8$ and $Q_{12}$ together depend on 22 complex parameters, after fixing 4 of them we are left with 18 adjustable complex parameters. Hence, the 24 positions of the 7-branes are parameterized in terms of 18 complex parameters. The absolute value of $c$ can be associated with a real K\"ahler modulus, while the 18 complex parameters can be associated with the complex structure moduli of the K3. The argument of $c$ can be absorbed into a redefinition of $\epsilon_0$ and does not represent a modulus.

\begin{figure}
\centering
\vskip -30cm
\psfrag{iinfty}{$i\infty$}
\psfrag{infty}{$\infty$}
\psfrag{0}{0}
\psfrag{1}{1}
\psfrag{i}{$i$}
\psfrag{jt}{$j(\tau)$}
\psfrag{jtz}{$j(\tau(z)) = \frac{P^3_8}{P^3_8 + Q^2_{12}}$}
\psfrag{F}{$F$}
\psfrag{rho}{$\rho$}
\psfrag{ziinfty}{$z_{i\infty}$}
\psfrag{zrho}{$z_\rho$}
\psfrag{zi}{$z_i$}
\psfrag{x24}{$\times 24$}
\psfrag{x8}{$\times8$}
\psfrag{x12}{$\times 12$}
\psfrag{tplane}{$\angle^\tau$}
\psfrag{jplane}{$\angle^j$}
\psfrag{zplane}{$\angle^z$}
\includegraphics{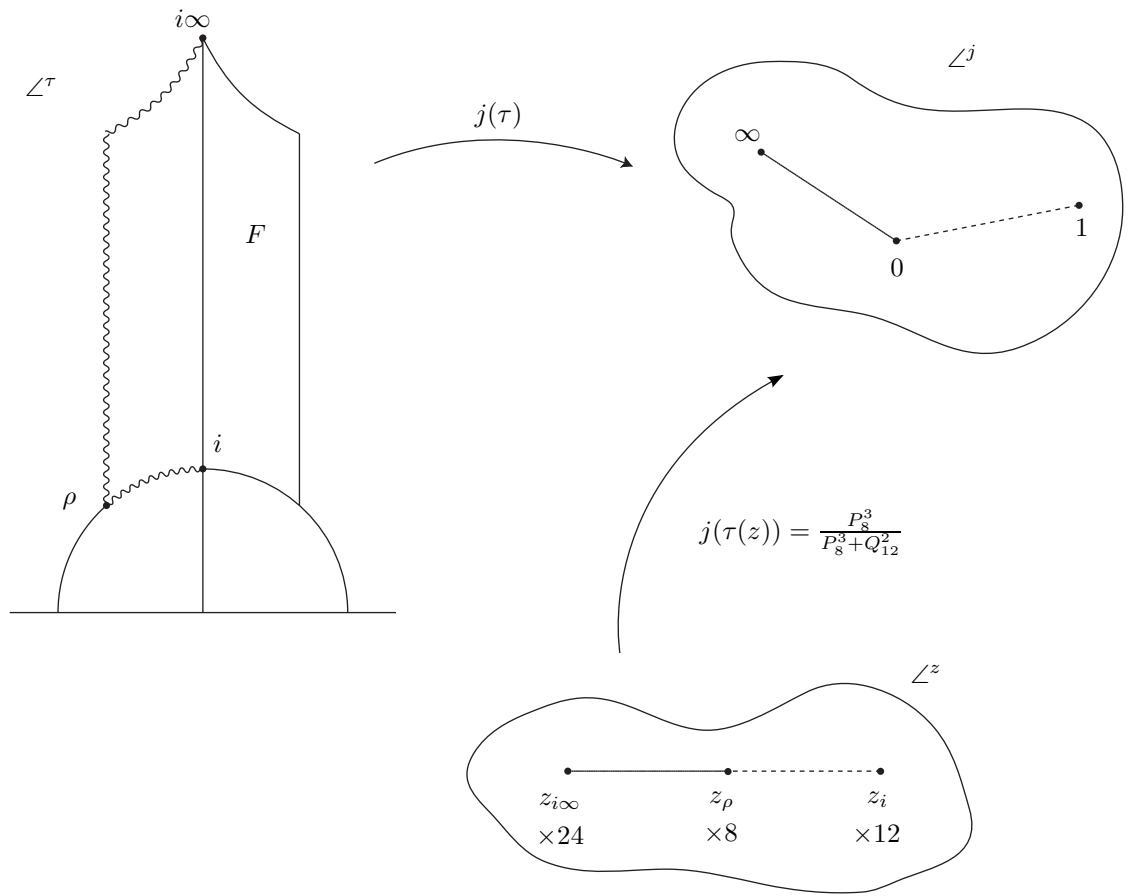}
\caption{In the top left figure we indicate our choice of fundamental domain of
the group $PSL(2,\mathbb{Z})$. The top right figure summarizes the
transformation properties of the $j$-function. The bottom figure is a
schematic representation of the branch cuts of the function
$\tau(z)$. The solid (dashed) line indicates a branch cut with a $T (S)$ transformation.}\label{fig:mappingproperties}
\end{figure}

The $z$-dependence of the axi-dilaton  $\tau$ is summarized in
Fig. \ref{fig:mappingproperties}. The top left figure indicates the chosen fundamental domain
$F$ of $PSL(2,\mathbb{Z})$, together with its orbifold points
$\tau=i\infty,\tau  = \rho$ and $ \tau = i$. The $j$-function maps
these orbifold points to the points $j=\infty, j=0$ and $j=1$,
respectively. The top right figure indicates the branch cuts of the
inverse $j$-function. The bottom figure
shows that the $j$-plane is mapped 24 times onto the z-plane. Under
this mapping the point $j=\infty$ is mapped to 24 distinct points
$z_{i\infty}$ which are the 24 zeros of the polynomial $P_8^3 +
Q_{12}^2$. Similarly, the points $j=0$ and $j=1$ are mapped to 8
distinct points $z_\rho$ (which are the 8 zeros of $P_8$) and 12
distinct points $z_i$ (which are the 12 zeros of $Q_{12}$),
respectively. The points $z_{i\infty}$, $z_\rho$ and $z_i$ are those points where $\tau$ takes the values
$i\infty$, $\rho$ and $i$, respectively\footnote{This definition applies to a situation in which $\tau$ does not take its values in the covering space, but always in the fundamental domain.}. The branch  cuts of the inverse $j$-function, i.e. of $\tau$ as a function of $z$, are indicated schematically in the lower figure of Fig. \ref{fig:mappingproperties}. The precise branch cut structure will be discussed in Subsection \ref{subsec:graphical}.

Going counterclockwise around the branch points $z_{i\infty}$ or $z_i$, we measure a  $T$ or $S$ $PSL(2,\mathbb{Z})$ transformation on $\tau$,
with  $T$ and $S$ given by
\begin{equation}T= \begin{pmatrix}
1&1\cr 0&1\end{pmatrix}\,,\hskip 2truecm
S= \begin{pmatrix}
0&1\cr -1&0\end{pmatrix}\,.
\end{equation}
We will indicate a branch cut with a $T(S)$ transfomation by a solid (dashed) line, as in Fig. \ref{fig:mappingproperties}. Later, we will refine this notation in order to also include the transformation of $f$ which transforms under $SL(2,\mathbb{Z})$ instead of $PSL(2,\mathbb{Z})$:
\begin{equation}\label{trafoSL2Z}
\tau\rightarrow\ \frac{a\tau+b}{c\tau+d}\,,\hskip 2truecm
f\rightarrow (c\tau+d)f\,.
\end{equation}

Sometimes we will use a description in which $\tau$ and $f$ take their values in the covering space, which for $\tau$ is the entire upper half plane. This means that when a branch cut is crossed $\tau$ and $f$ continuously change their values from one branch into an adjacent one. Sometimes we will use a description in which $\tau$ and $f$ take their values on a particular branch which for $\tau$ is a fundamental domain. In that case when a branch cut is crossed $\tau$ and $f$ change their values discontinuously.

\subsection{Singularities and Gauge Groups}\label{subsec:ADEsingularities}

The 24 non-coincident 7-branes are located at the points
$z_{i\infty}$. At each of these points a 1-cycle of the fibered 2-torus shrinks to a point.
When a number of such points is made to
coincide different types of singularities are formed. The type of
singularity depends on the details of the zeros of $P_8^3+Q_{12}^2$,
i.e. whether or not the zero of $P_8^3+Q_{12}^2$ is also a zero of
either $P_8$ and/or $Q_{12}$ and what the orders of the zeros of
$P_8$, $Q_{12}$ and $P_8^3+Q_{12}^2$ are. The singularities of an
elliptically fibered 2-torus have been classified by Kodaira (see
for example \cite{Barth}) and the relation between the singularity
type of the singular fibre with the order of the zeros of $P_8$,
$Q_{12}$ and $P_8^3+Q_{12}^2$ follows from applying Tate's algorithm
\cite{Tate}. The possible singularities are listed in Table
\ref{Tatealgorithm} which has been adopted from
\cite{Bershadsky:1996nh}.

Table \ref{Tatealgorithm} is useful in determining the non-Abelian
factors of the 7-brane gauge groups. An $A_{n-1}$ singularity for
$n\ge 2$ leads to a gauge group $SU(n)$, a $D_{n+4}$ singularity to
a gauge group $SO(2(n+4))$ and the $E_6$, $E_7$ and $E_8$
singularities lead to the exceptional gauge groups $E_6$, $E_7$ and
$E_8$. The third to fifth rows of Table \ref{Tatealgorithm}
correspond to the Argyres--Douglas singularities
\cite{Argyres:1995jj,Argyres:1995xn}. The singularity type indicated by ``none'' in the first row
of Table \ref{Tatealgorithm}  refers to the fact that the 7-brane
gauge group is trivial (there is no 7-brane since the order of the zero of $P_8^3+Q_{12}^2$ is zero), while the same singularity type in the third row means that the gauge group is Abelian.

Besides the non-Abelian gauge groups predicted by Table \ref{Tatealgorithm} there can additionally be various $U(1)$ factors coming from the 7-branes. We recall, as discussed in Subsection \ref{subsec:solutions}, that the solution has 18 free complex parameters, which can be associated with the complex structure moduli of the K3. From the 8-dimensional point of view these complex moduli reside in minimal vector supermultiplets. Therefore there are 18 $U(1)$ factors for 24 7-branes\footnote{We remark that there are additionally two more vectors residing in the minimal 8-dimensional gravity supermultiplet. These vectors do not participate in the symmetry enhancement.}. The fact that there is not a one-to-one relationship between $U(1)$ factors and 7-branes is because there are certain global obstructions to the positioning of the 24 7-branes. This will be further discussed in Subsection \ref{subsection:charges}.

\begin{table}[ht!]
\begin{center}
\begin{tabular}{|c|c|c|c|c|}
  \hline
  Order zero  & Order zero  & Order zero  & Singularity  & Conjugacy \\
  $P_8$ & $Q_{12}$ & $P_8^3+Q_{12}^2$ & Type & class \\
  \hline
  &&&&\\
    $\ge 0$ & $\ge 0$ & $0$ & none & $[\mathbbm{1}]$\\
  &&&&\\
    $0$ & $0$ & $n$ & $A_{n-1}$ & $[T^n]$\\
  &&&&\\
    $\ge 1$ & $1$ & $2$ & none  & $[T^{-1}S]$ \\
  &&&&\\
    $1$ & $\ge 2$ & $3$ & $A_1$  & $[S]$ \\
  &&&&\\
    $\ge $2 & $2$ & $4$ & $A_2$  & $[(T^{-1}S)^2]$\\
  &&&&\\
    $2$ & $\ge 3$ & $n+6$ & $D_{n+4}$ & $[-T^n]$\\
  &&&&\\
    $\ge 2$ & $3$ & $n+6$ & $D_{n+4}$ & $[-T^n]$\\
  &&&&\\
    $\ge 3$ & $4$ & $8$ & $E_6$ & $[-T^{-1}S]$\\
  &&&&\\
    $3$ & $\ge 5$ & $9$ & $E_7$ & $[-S]$\\
  &&&&\\
    $\ge 4$ & $5$ & $10$ & $E_8$ & $[-(T^{-1}S)^2]$\\
  &&&&\\
  \hline
\end{tabular}
\end{center}
\caption{{\small{The Kodaira classification of singular fibres of an
elliptically fibered 2-torus.  When the singularity in the fourth column is
called `none' it means that the group contains no non-Abelian
factor. The last column indicates the $SL(2,\mathbb{Z})$ conjugacy class of the singularity.}}}
\label{Tatealgorithm}
\end{table}

We now consider the local geometry of the 7-brane solution in the
neighborhood of a singularity. Consider for example the $E_6$
singularity for which the order of the zero of $Q_{12}$ must be
four. Suppose that this singularity is located at the point $z_0$ in the
transverse space. Then near $z_0$ we have
$Q_{12}\approx \rm{cst}(z-z_0)^4$. Further, according to Table
\ref{Tatealgorithm} we have $P_8 \approx \rm{cst}(z-z_0)^n$ with $n\ge 3$
and $P_8^3+Q_{12}^2 \approx \rm{cst}(z-z_0)^{8}$ so that near $z_0$ we have
$f \approx \rm{cst}\,\eta^2(\tau)(z-z_0)^{-2/3}$ and
$j(\tau)\approx\rm{cst}(z-z_0)^{3n-8}$. Since $n\ge 3$ we have $j=0$ at $z=z_0$, i.e. $\tau=\Lambda\rho$ where $\Lambda$ is some $PSL(2,\mathbb{Z})$ transformation. The most general transformation
that $\tau$ may undergo compatible with this local
expression is of the form
$\Lambda(\pm T^{-1}S)\Lambda^{-1}$. Next, the sign can be fixed by comparing the transformation of $f$ with Eq. \eqref{trafoSL2Z}. Thus we conclude that $E_6$
singularities are formed at fixed points of $SL(2,\mathbb{Z})$
transformations that belong to the $-T^{-1}S$ $SL(2,\mathbb{Z})$
conjugacy class, that we will denote by $[-T^{-1}S]$.
In a similar way one can relate the other singularities to $SL(2,\mathbb{Z})$ conjugacy classes. These conjugacy classes are indicated in the last column of Table \ref{Tatealgorithm}.

\subsection{Graphical Representations}\label{subsec:graphical}

\begin{figure}
\centering
\psfrag{ziinfty}{$z_{i\infty}$}
\psfrag{zi}{$z_i$}
\psfrag{T}{$T$}
\psfrag{S}{$S$}
\psfrag{minS}{$-S$}
\psfrag{min1}{$-1$}
\includegraphics{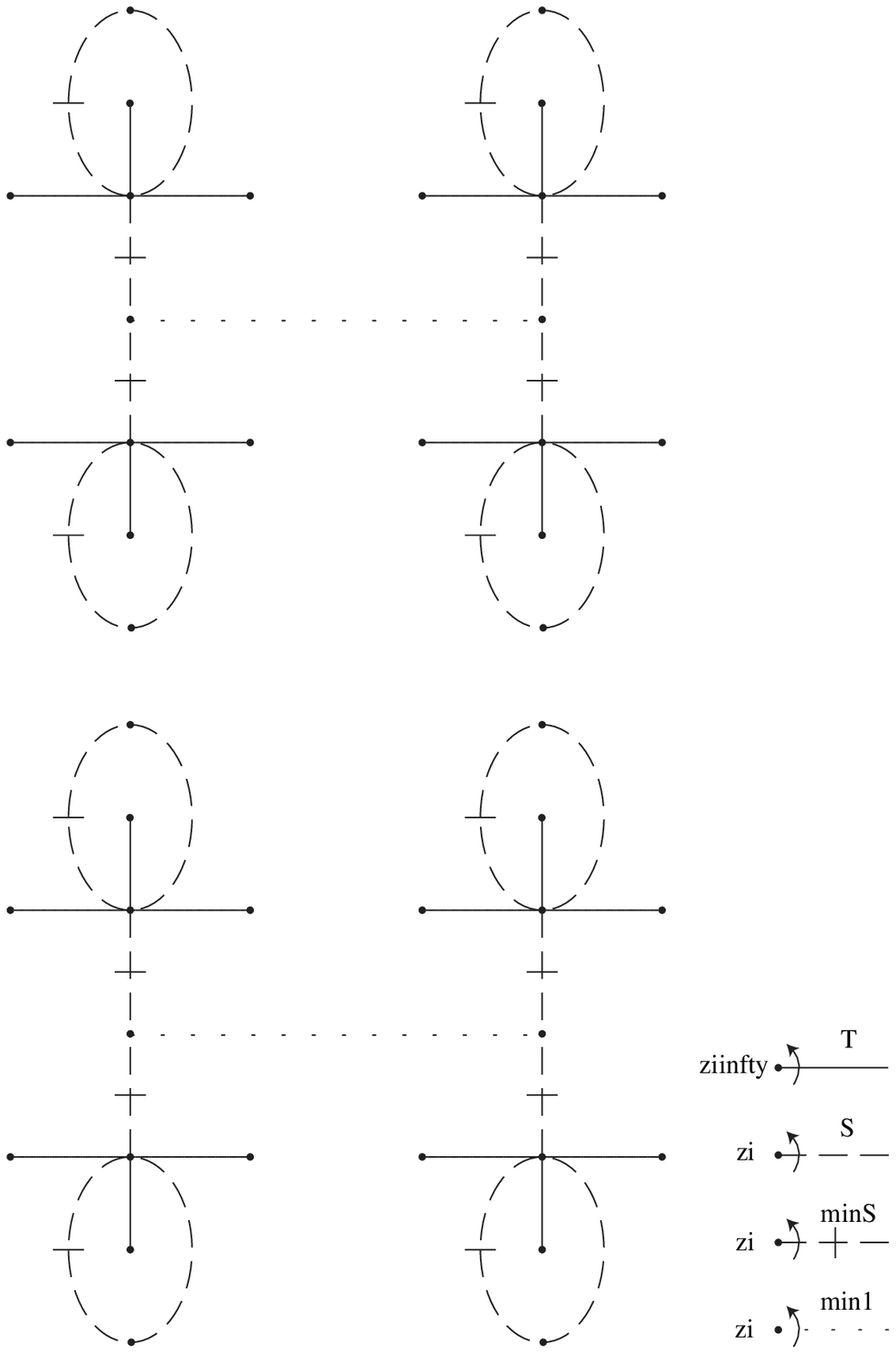}
\caption{This figure shows a particular representation of the branch cuts of the pair
$(\tau(z),f(z))$. The $T$ transformations are measured when crossing
a solid branch cut counterclockwise around $z_{i\infty}$. The $S$,
$-S$ and $-\mathbbm{1}$ transformations are measured when crossing
the respective dashed, dashed with a minus sign through and dotted
branch cuts counterclockwise around the branch points $z_i$.}\label{fig:branchcutstructure}
\end{figure}

We first discuss a refinement of Fig. \ref{fig:mappingproperties}, which also takes into
account  the transformation properties of $f$. Our starting point is
the top right figure in Fig. \ref{fig:mappingproperties}. We will first distill out of this
figure a new Fig. \ref{fig:branchcutstructure} which contains the detailed branch cut
structure of the function $\tau(z)$. This new figure can be obtained
as follows. We start by taking  all the points
$z_{\rho}^1,\ldots,z_{\rho}^8$, $z_{i}^1,\ldots, z_i^{12}$ and
$z_{i\infty}^1,\ldots,z_{i\infty}^{24}$ non-coinciding. The inverse
$j$-function has branch cuts running from $j=\infty$ to $j=0$ and
from $j=0$ to $j=1$. This means that each point $z_{i\infty}^1$ to
$z_{i\infty}^{24}$ has a branch cut connecting it to one of the eight points $z_{\rho}^1,\ldots,z_{\rho}^8$. At each point $z_{\rho}$ we thus have three $T$ branch cuts. Next we must include the $S$ branch cuts that connect the points $z_{\rho}$ and $z_i$. Since there are three $T$ branch cuts meeting at $z_{\rho}$ we must have three $S$ branch cuts meeting at $z_{\rho}$ as well. These eight times three $S$ branch cuts must then end on twelve points $z_{i}$. At each point $z_i$  two $S$ branch cuts meet. Now, we need to put the $T$ and $S$ branch cuts that meet at $z_{\rho}$ in such an order that the monodromy around $z_{\rho}$ is the identity in $PSL(2,\mathbb{Z})$. This will be the case if
we put the three $T$ and three $S$ branch cuts meeting at $z_\rho$ in alternating order. This follows from the $SL(2,\mathbb{Z})$ identity $(TS)^3=\mathbbm{1}$. Next we need to find the positioning of the $S$ branch cuts. At this point there are two choices:
\begin{enumerate}
 \item We take two of the three $S$ branch cuts that originate from the same branch point $z_\rho$ to go to the 
{\it same} point $z_i$.
\item We take all three $S$ branch cuts to go to three {\it different} points $z_i$.
\end{enumerate}
This shows that our construction of the graphical representation is not unique. If we always choose the first option then we obtain the $PSL(2,\mathbb{Z})$ part of Fig. \ref{fig:branchcutstructure}. The minus signs through some of the $S$ branch cuts refer to the $SL(2,\mathbb{Z})$ properties of the figure as will be explained next.

Now that we have fixed the positioning of the $S$ branch cuts, we
can consider the transformation properties of the function $f$. From
the expression for $f$, Eq. \eqref{IIBbackgroundf}, we know that $f$
does not transform when going locally around any of the points
$z_{i\infty}$, $z_\rho$ or $z_i$. Let us first consider those points
$z_i$ at which two $S$ branch cuts meet that come from the same
point $z_\rho$. Since $S^{2}=-\mathbbm{1}$ and the function $f$ transforms under the
$-\mathbbm{1}$ element of $SL(2,\mathbb{Z})$ we need to take one of
these two $S$ branch cuts to be a $-S$ branch cut. This has no
effect on the transformation properties of $\tau$ and realizes that
$f\rightarrow f$ when going around these $z_i$ points. There are
eight $z_i$ points in Fig. \ref{fig:branchcutstructure} which can be treated in this way. Consider now the
points $z_\rho$. In order that $f\rightarrow f$ when going around
$z_\rho$ we need to turn the $S$ branch cut that goes to a point
$z_i$ at which it meets an $S$ branch cut coming from another point
$z_\rho$ into a $-S$ branch cut, so that going counterclockwise
around $z_\rho$, not encircling any other branch points, gives the
$+\mathbbm{1}$ element of $SL(2,\mathbb{Z})$. As a result we are now
left with four points $z_i$ at which two $-S$ branch cuts meet. The
only way to realize that $f\rightarrow f$ when going around any of
these four points $z_i$ is to introduce a new branch cut. This
new branch cut cannot have any effect on $\tau$, so it must be a
$-\mathbbm{1}$ branch cut. Further, because the only points around
which $f$ does not yet transform to itself are these four points $z_i$, the
$-\mathbbm{1}$ branch cut must go from one point $z_i$ to another
point $z_i$. Since we have four points $z_i$ at which a
$-\mathbbm{1}$ branch cut ends we need two such new branch cuts.
Finally, in order not to introduce additional branch points these
$-\mathbbm{1}$ branch cuts are not allowed to intersect each
other. This explains all the
necessary steps to construct Fig. \ref{fig:branchcutstructure}. This figure provides a representation of the branch
cuts of the pair $(\tau(z),f(z))$.
Note that the way in which we have chosen to place the $-S$ and
$-\mathbbm{1}$ branch cuts is not unique. We could, for example, not
use any $-S$ branch cuts and instead place a $-\mathbbm{1}$ branch
cut between pairs of points $z_i$ in such a way that the
$-\mathbbm{1}$ branch cuts do not intersect.

It is useful to highlite all the choices that went into Fig. \ref{fig:branchcutstructure}.
These choices have been:
\begin{enumerate}
    \item Choose a fundamental domain.
    \item Choose a positioning of the $S$ branch cuts that leads to a proper representation of
    $\tau(z)$.
    \item Choose a positioning of the $-S$ and $-\mathbbm{1}$ branch
    cuts that also takes into account the transformation properties of $f(z)$.
\end{enumerate}
Even though the branch cut representations are not unique the allowed choices for the positioning of the $S$ branch cuts lead to certain global obstructions. We note that for each choice of positioning of the $S$ branch cuts it is possible to put certain 7-branes on top of each other (without modifying the $S$ branch cuts). This leads to manifest $SU$ symmetry groups. It will be shown later that different positionings of the $S$ branch cuts correspond to different embeddings of these $SU$ groups into the to be formed $A$-$D$-$E$-gauge groups.

\subsection{Charges}\label{subsection:charges}

The charges of a 7-brane located at a point $z_{i\infty}$ will be
meausured around an infinitesimal loop encircling the point
$z_{i\infty}$. A single 7-brane has charges $p$ and $q$ when
the $\tau$-monodromy along an infinitesimal loop around $z_{i\infty}$ is of the form:
\begin{equation}\label{pqcharges}
\Lambda T\Lambda^{-1} \hskip 2truecm \text{with} \hskip 2truecm
\Lambda=\begin{pmatrix} p&r\cr q&s\end{pmatrix}\,,
\end{equation}
and $sp-qr=1$. With this definition each point $z_{i\infty}$
corresponds to a $(1,0)$ 7-brane due to our choice of fundamental
domain. However, there are some $(1,0)$ 7-branes that are encircled
by $\pm S$ branch cuts. Thus, from the point of view of a base point
that lies outside the region enclosed by two $S$ branch cuts these
7-branes appear to have different charges $p$ and $q$ depending on
the loop that one uses to encircle such $S$ branch cut locked $(1,0)$ 7-branes. From now on we will refer to such $S$ branch cut locked 7-branes simply as locked 7-branes.

The monodromy loops giving rise to different $p$
and $q$ charges are always large\footnote{For example, in the orientifold limit introduced by Sen \cite{Sen:1997gv} the monodromy around the two 7-branes that describe the split orientifold is computed at a finite distance from the two branes.} and never infinitesimally close to
the locked points $z_{i\infty}$. From this point of view one cannot view the locked 7-brane as being some $(p,q)$ 7-brane. For instance, although one can always put two $(1,0)$ 7-branes on top of each other one cannot force two locked 7-branes to coincide without altering the branch cut structure. The infinitesimal monodromy around a 7-brane
is determined by the choice of fundamental domain and this
can be chosen only once. If we denote the fundamental domain displayed
in the top left figure of Fig. \ref{fig:mappingproperties} by $F$ then the fundamental domain
$\Lambda[F]$ contains $(p,q)$ 7-branes as
defined in \eqref{pqcharges}.

In the graphical representation of Fig. \ref{fig:branchcutstructure} there are 16 $(1,0)$
7-branes that are not locked by $S$ branch cuts. This number depends on the graphical representation.
For instance, another choice of positioning the
$S$ branch cuts exists that leads to 18
$(1,0)$ 7-branes that are not locked by $S$ branch cuts. This branch cut representation can be obtained by taking twice the upper figure of Fig. \ref{fig:alternatives} in which the five coinciding 7-branes are taken apart. It is not possible to position the $S$ branch cuts in such a way that we have more than 18 unlocked branes. This agrees with the earlier observation that the maximal rank of the gauge group is 18. In the orbifold limits of $K3$ this can be observed from the analysis of \cite{Dasgupta:1996ij}.

\subsection{Other 7-brane solutions}\label{subsec:othersolutions}

We end this section with a few comments about other 7-brane
solutions. The non-compact solutions with 6 or 12 7-branes have a
graphical representation that can be inferred from Fig. \ref{fig:mappingproperties} by
simply considering the 24 7-brane solutions as four copies of a
solution with 6 or 2 copies of a solution with 12 7-branes. Further,
it can be shown that any other supersymmetric 7-brane solution with
$\tau(z=\infty)$ arbitrary can be formed out of the 6, 12 or 24 7-brane solutions by taking certain 7-branes to coincide. This includes both solutions containing so-called Q7-branes as well as
solutions for which the $\tau$ monodromy group is a subgroup of
$PSL(2,\mathbb{Z})$. An example of the latter kind is given in \cite{Bergshoeff:2006jj}.
The name Q7-brane has been coined in \cite{Bergshoeff:2007aa} to
refer to a solution with $\tau(z)$ non-constant that contains
deficit angles at the points $z_{\rho}$ and $z_i$. Such deficit angles arise
when a locked 7-brane is put on top of an unlocked one\footnote{For a world-volume discussion supporting the point of view that a Q7-brane corresponds to a stack of 7-branes in which some 7-branes have charges that are $PSL(2,\mathbb{Z})$ transformed with respect to other 7-branes in the stack, see \cite{Bandos:2008bn}.}. The global branch cut structure of solutions with either Q7-branes or a monodromy group that is a subgroup of $PSL(2,\mathbb{Z})$ cannot be obtained by continuously deforming the $T$ and $S$ branch cuts of a solution with 24 non-coinciding 7-branes and must therefore be studied separately using new branch cut rules.

The result of putting a locked 7-brane on top of an unlocked one in the supergravity approximation appears as a single brane that couples to an 8-form potential that is outside the $SL(2,\mathbb{Z})$ orbit of the RR 8-form. By electro-magnetic duality this same 8-form potential magnetically sources so-called Q-instantons \cite{Bergshoeff:2008qq}. The Q-instantons relate to new vacua of the quantum axi-dilaton moduli space $SO(2) \backslash PSL(2,\mathbb{R})\slash PSL(2,\mathbb{Z})$ and are argued to be relevant for the IIB theory in the neighborhood of the orbifold points $\tau_0=i,\rho$ (and their $PSL(2,\mathbb{Z})$ transforms) of the quantum moduli space.

\section{BPS open strings}\label{sec:BPSstrings}

Consider any of the 24 7-branes of Fig. \ref{fig:branchcutstructure} and consider an open $(1,0)$ string that has not yet crossed any branch cuts and that has one endpoint ending on the 7-brane which is thus a $(1,0)$ 7-brane. Let us now follow this string along some path $\gamma$ that will generically cross some number of branch cuts going from this 7-brane to another one.

If we allow $\tau$ (and $f$) to take values in the covering space then when a branch cut is crossed $\tau$ changes its values continuously from one branch (fundamental domain) into an adjacent one.
The string tension of a $(1,0)$ string is then only continuous across the branch cut if the string charges do not transform, i.e. a $(1,0)$ string remains a $(1,0)$ string. If the $(1,0)$ string would cross a number of branch cuts whose overall $SL(2,\mathbb{Z})$ transformation equals $\Lambda$ and subsequently approaches a 7-brane that 7-brane would appear to be some $(p,q)$ 7-brane with monodromy $\Lambda T\Lambda^{-1}$, i.e.~the 7-brane charges have changed. Alternatively, we can assume that $\tau$ and $f$ always take their values on some particular branch. In this case $\tau$ and $f$ change their values discontinuously when crossing a branch cut. If we do this then all the 7-branes are always of $(1,0)$ type. In order for the string tension to change continuously when the string crosses a number of branch cuts whose overall $SL(2,\mathbb{Z})$ transformation is $\Lambda$ the string charges at the end of the string must be $\Lambda\textstyle{\left(\begin{array}{c}1\\0
\end{array}\right)}$. In summary, from the point of view of a string, working in the covering space means that the 7-brane charges transform, while working with a fixed branch means that the string charges transform.

In any case, regardless one's point of view, going back to the $(1,0)$ string with one endpoint on the 7-brane, in order for it to end on another 7-brane it must cross a number of branch cuts for which $\Lambda$ is such that it leaves the $(1,0)$ string invariant, i.e. $\Lambda=\pm T^k$ for some $k\in\mathbb{Z}$.

Paths $\gamma$ along which the overall $SL(2,\mathbb{Z})$ transformation is $\pm T^k$ can in general have self-intersections. However, if these paths are to represent possible profiles of open strings self-intersections are not allowed. Classically, this is because the endpoints of an open string move at the speed of light in order to counteract the tension of the string preventing it from collapsing to a point. There is no way to sustain a closed loop that would be formed if the string were to self-intersect. Therefore we must restrict to simple paths, i.e.~paths without self-intersections, along which the overall $SL(2,\mathbb{Z})$ transformation is given by $\Lambda=\pm T^k$.

We can always have the $(1,0)$ string loop a sufficient number of times around the begin or end brane, or when a number of 7-branes coincide at one point have the $(1,0)$ string to cross, in a suitable manner, a sufficient number of $T$ branch cuts, so that the overall $SL(2,\mathbb{Z})$ transformation along the string is $\pm\mathbbm{1}$. As will be explained in the next section, this does not lead to inequivalent strings. Therefore, from now on we will restrict our attention to those simple curves along which $\Lambda=\pm\mathbbm{1}$.

A string stretched between two non-coinciding 7-branes would become massless if it must always lie along a simple path connecting the two 7-branes. Hence, in order to understand the open string origin of 7-brane gauge groups we must find those strings that lie along simple curves along which $\Lambda=\pm\mathbbm{1}$ whose masses are BPS. Once these BPS strings have been identified one can try to associate them to various generators of the 7-brane gauge group.

We will assign an orientation to each open string. When $\Lambda=-\mathbbm{1}$ along the BPS string then with $\tau, f$ taking values in the covering space the $(1,0)$ string starts on a $(1,0)$ 7-brane and ends on a $(-1,0)$ 7-brane. Or with $\tau, f$ taking their values on some fixed branch the string starts as a $(1,0)$ string on a $(1,0)$ 7-brane and ends as a $(-1,0)$ string on a $(1,0)$ 7-brane. In the latter case the directionality along the string has flipped.

Due to our choice of fundamental domain a string always starts and ends on a $(1,0)$ 7-brane (up to signs). If we allow $\tau$ and $f$ to take values in the covering space the string charges do not transform and it is sufficiently general to consider only the tension of a $(1,0)$ string in order to compute the mass of the stretched string. The mass of a $(1,0)$
string that is stretched along some simple curve $\gamma$, denoted by $m_{1,0}$, is given by
\begin{equation}
    m_{1,0}=\int_{\gamma}T_{1,0}ds=\int_{\gamma}\vert fdz\vert=\int_{\gamma}\vert dw_{1,0}\vert\,,
\end{equation}
where $ds$ is the Einstein frame line element, $T_{1,0}$ is
the tension of a $(1,0)$ string,
\begin{equation}
    T_{1,0}=\left(\text{Im}\,\tau\right)^{-1/2}\,,
\end{equation}
and $dw_{1,0}$ is defined to be
\begin{equation}
    dw_{1,0}=fdz\,.
\end{equation}
Denoting the path $\gamma$ for which the mass of the $(1,0)$ string satisfies
a lower bound by $\gamma_{\text{BPS}}$, the BPS mass is given by
\begin{equation}\label{BPS}
    m_{1,0}^{\text{BPS}}=\vert\int_{\gamma_{\text{BPS}}}dw_{1,0}\vert\,.
\end{equation}
Hence BPS strings lie along $\gamma_{\text{BPS}}$. These are non self-intersecting paths from $z^i_{i\infty}$ to $z_{i\infty}^j$ with $i\neq j$ and $i,j=1,\ldots,24$ along which the overall $SL(2,\mathbb{Z})$ transformation is $\pm\mathbbm{1}$. This definition applies to a situation in which none of the 7-branes are coinciding. As we will see when some 7-branes are put on top of each other at, say, $z_{i\infty}^1$ then there can also exist BPS strings that lie along paths which are non-self-intersecting along which the overall $SL(2,\mathbb{Z})$ transformation is $-\mathbbm{1}$ that go from $z_{i\infty}^1$ back to itself along some non-contractible loop.

\section{Open strings and the $A$-$D$-$E$-singularities}\label{sec:gaugegroups}

Before we discuss specific cases we first make some general observations.
In order to study symmetry enhancement using open strings we need to isolate those 7-branes which when made to coincide give rise to a certain gauge group. These branes can be identified as follows.
In the solution with no 7-branes coinciding the branes that when made to coincide give rise to the gauge groups of Table
\ref{Tatealgorithm} can be found by encircling a group of 7-branes that satisfy two criteria:
\begin{enumerate}
\item The number of 7-branes that is encircled is given by the third column of Table 1.
\item When encircling these 7-branes by a loop (with
winding number one) the $SL(2,\mathbb{Z})$ monodromy must belong to the
conjugacy class\footnote{This loop can be continuously deformed to any other loop that encircles the same number of branes and that belongs to the same $SL(2,\mathbb{Z})$ conjugacy class.} that is given in the fifth column of Table 1.
\end{enumerate}
Examples of monodromy loops encircling a group of 7-branes which when made to coincide give rise to certain $A$- and $D$-type gauge groups are given in Fig. \ref{fig:conjugacyclassloops}.

One could also encircle a group of 7-branes that does not satisfy the above two criteria. For those cases there is no limit in which the group of 7-branes can be made to coincide, so that BPS strings cannot become massless. Such groups of non-collapsable 7-branes have been studied in \cite{DeWolfe:1998eu,DeWolfe:1998pr} and will not be considered further here.

The loop encircling a number of 7-branes around which the monodromy belongs to some $SL(2,\mathbb{Z})$ conjugacy class can be considered to form the boundary of a punctured disk where each puncture corresponds to some 7-brane. In general, a necessary condition for two strings to be inequivalent, is that they lie along two non-selfintersecting $\Lambda=\pm\mathbbm{1}$ paths that are homotopically distinct in the sense of the homotopy of this punctured disk. There is an important exception to this statement which concerns strings crossing branch cuts of branes they can end on. A $(1,0)$ string going from point $A$ to point $B$ (both lying inside the disk) along $\gamma_a$ crossing a $T$ branch cut (and no other branch cuts) is equivalent to a $(1,0)$ string going from point $A$ to point $B$ along $\gamma_b$ without crossing any branch cuts, i.e.~the string along $\gamma_b^{-1}\gamma_a$ is contractible. Put another way, for strings that only cross $T$ branch cuts and nothing else the inequivalence of strings only depends on the starting point and endpoint and not on the homotopy.

Before we embark on a discussion of open BPS strings and gauge group generators let us briefly recall the symmetry enhancement for a stack of $n$ D7-branes in perturbative string theory. In this case the symmetry group is $U(n)$. The Cartan subalgebra of $U(n)$ which is $(U(1))^n$ has as many $U(1)$ factors as there are D7-branes. Each string has a definite orientation and the charges at the string endpoints couple to vectors that are associated to the $U(1)$ of the Cartan subalgebra. The symmetry enhancement comes from massive BPS strings stretched between the different D7-branes. Taking all possible orientations into account there are $n(n-1)$ such BPS strings \cite{Witten:1995im}. In the context of perturbative string theory this analysis has been generalized to include orientifolds \cite{Gimon:1996rq}. For example consider a stack of 4 D7-branes and one O7-plane\footnote{In the context of the global 7-brane solution the O7-plane can be viewed as an approximate solution in which the two locked 7-branes of Fig. \ref{fig:branchcutssu(4)} have been put on top of each other, see \cite{Sen:1996vd}. An exact solution with an O7-plane is obtained only once the four D7-branes are put on top of the O7-plane.}. In this case the gauge group is $SO(8)$. The 4 D7-branes give rise to the gauge group $U(4)$. By including additional BPS strings that go from the stack of 4 D7-branes to the O7-plane and back additonal Chan--Paton states originate that together with the $U(4)$ states give rise to $SO(8)$. The Cartan subalgebra of $SO(8)$ is given by the Cartan subalgebra of $U(4)$.

This familiar situation from perturbative string theory does not straightforwardly apply to the case of F-theory on K3. This is for a number of reasons. First of all we note that the number of 7-branes in the third column of Table 1 does not equal the rank of the gauge groups. For $A_{n-1}$, $D_{n+4}$, $E_6$. $E_7$ and $E_8$ we have
\begin{equation}\label{ranknumberbranes}
\text{\#7-branes}=\text{rank of non-Abelian part of 7-brane gauge group}+m\,,
\end{equation}
where $m=1$ for the A series related to the $[T^n]$ conjugacy classes and $m=2$ for the $A$ series related to the Argyres--Douglas singularities as well as for the $D$ and $E$ series. Hence, the Cartan subalgebra is not related to the individual branes. Further, as discussed earlier, the total number of $U(1)$ factors coming from 7-branes is 18 while the total number of 7-branes is 24.

Another complication that follows from \eqref{ranknumberbranes} is that, except for the $[T^n]$ conjugacy classes, the number of BPS strings is larger than the number of generators in the gauge group that lie outside the Cartan subalgebra. One can draw BPS strings between each pair of 7-branes inside the loop that encircles the 7-branes forming the singularity. Further, one can draw homotopically inequivalent BPS strings between the same pair of 7-branes. We cannot umambigously say when the different BPS strings represent different generators because we have no means to assign the $U(1)$ charges to the string endpoints. We conclude that the relation between BPS strings and gauge group generators for the cases in which no branes are coinciding is not one-to-one.

The situation is improved when a subset of the 7-branes are made to coincide. This has two effects. First of all it enables us to use irreps of the non-Abelian algebra that is formed by putting some branes on top of each other to label the Chan--Paton states at the endpoints of the string. We can thus associate Chan--Paton states at the endpoints of the open strings with the definite location of a stack of some number of 7-branes. Secondly, the number of BPS strings will be much less as compared to a situation in which all branes are non-coincident\footnote{The reduction in the number of BPS strings is larger than the number of BPS strings that disappear due to the symmetry enhancement on the stack of branes.}. This latter fact can be explained as follows. Suppose two 7-branes are made to coincide along some path $\gamma$. Then any BPS string which crosses $\gamma$ seizes to exist once these two 7-branes are coincident. In fact as will be shown in the $E_6$ case putting a sufficiently large number of branes coincident can even lead to a situation in which no BPS strings exist at all. The effect of putting some number of 7-branes on top of each other is to make certain subgroups of the gauge group manifest.

Using Chan--Paton states for string endpoints on a stack of 7-branes only solves the problem of relating open strings to gauge group generators partially. Only those strings that have both their endpoints on a stack, which we will refer to as stack-to-stack-strings, can be related to gauge group generators. Since the $U(1)$ charges of the Cartan subalgebra are not manifest it is not possible to
tell when two strings ending on at least one single brane are (in)equivalent, so that we cannot map each individual string to some generator. Such strings occur in two types: 1). strings with both their endpoints on a different single brane are referred to as single-to-single-brane-strings and 2). strings with one endpoint on a stack and one endpoint on a single brane are referred to as stack-to-single-brane-strings. Below we will relate the stack-to-stack-strings in number to certain gauge group generators and for the stack-to-single-brane- and single-to-single-brane-strings we will derive consistency conditions in order for them to describe the remaining gauge group generators.

\subsection{$A$ series}\label{Aseries}

As follows from Table \ref{Tatealgorithm} the $A$-type gauge groups
are related to the following $SL(2,\mathbb{Z})$ conjugacy classes:
$[T^n]$, $[T^{-1}S]$, $[S]$ and $[(T^{-1}S)^2]$. For $[T^{-1}S]$ the gauge group
is Abelian.

The case of $[T^n]$ corresponds to the familiar situation of $n$ coiniciding D7-branes. We know from \cite{Witten:1995im} that the gauge group in this case is $U(n)$. We cannot however state that the $U(1)$ factor in $U(n)=U(1)\times SU(n)$ corresponds to this group of branes for reasons explained above.

\begin{figure}
\centering
\psfrag{minTb}{$[-T]$}
\psfrag{Sb}{$[S]$}
\psfrag{TinvSb}{$[T^{-1}S]$}
\psfrag{TinvS2b}{$[(T^{-1}S)^2]$}
\includegraphics{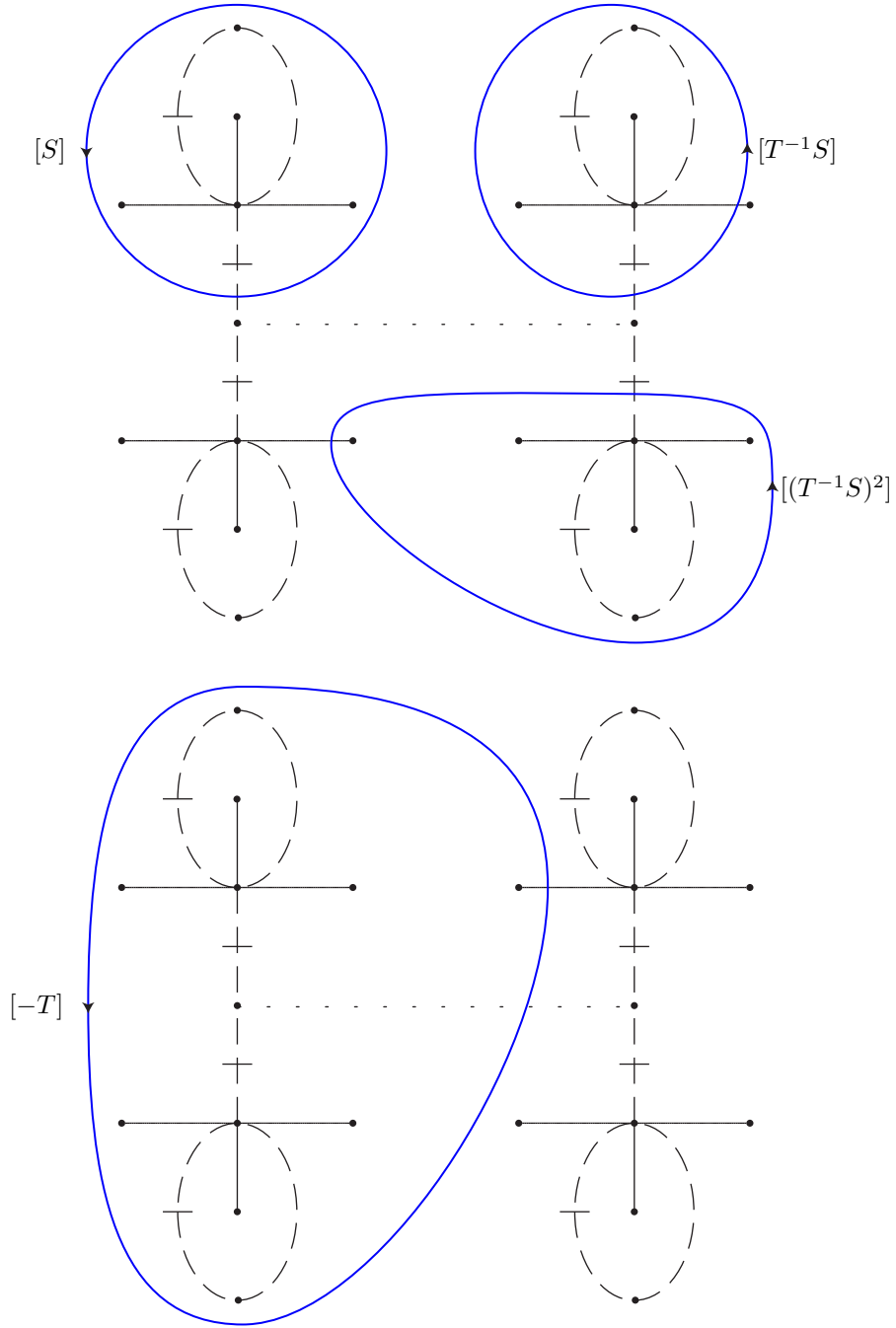}
\caption{This figure indicates a few examples of  loops that encircle a set of branes which when
made to coincide form the $A$- and $D$-type gauge groups. The corresponding conjugacy class (see the fifth column of Table 1)
is indicated.}\label{fig:conjugacyclassloops}
\end{figure}

The $SL(2,\mathbb{Z})$ conjugacy classes $[T^{-1}S]$, $[S]$ and $[(T^{-1}S)^2]$ correspond to the Argyres--Douglas singularities \cite{Argyres:1995jj,Argyres:1995xn}. The groups of 7-branes for these cases are shown in Fig. \ref{fig:conjugacyclassloops}.

For the $[T^{-1}S]$ case there does not exist a BPS string that lies inside the region bounded by the $[T^{-1}S]$ loop. Hence, no symmetry enhancement can occur. From Table \ref{Tatealgorithm} we know that indeed the gauge group is Abelian.

\begin{figure}
\centering
\psfrag{T1}{$T$}
\psfrag{T2}{$T$}
\psfrag{T3}{$T$}
\psfrag{T4}{$T$}
\psfrag{T5}{$T$}
\psfrag{[S]}{$[S]$}
\psfrag{[TinvSsquared]}{$[(T^{-1}S)^2]$}
\psfrag{I}{$1$}
\psfrag{II}{$2$}
\psfrag{1}{$1$}
\psfrag{2}{$2$}
\includegraphics[scale=.8]{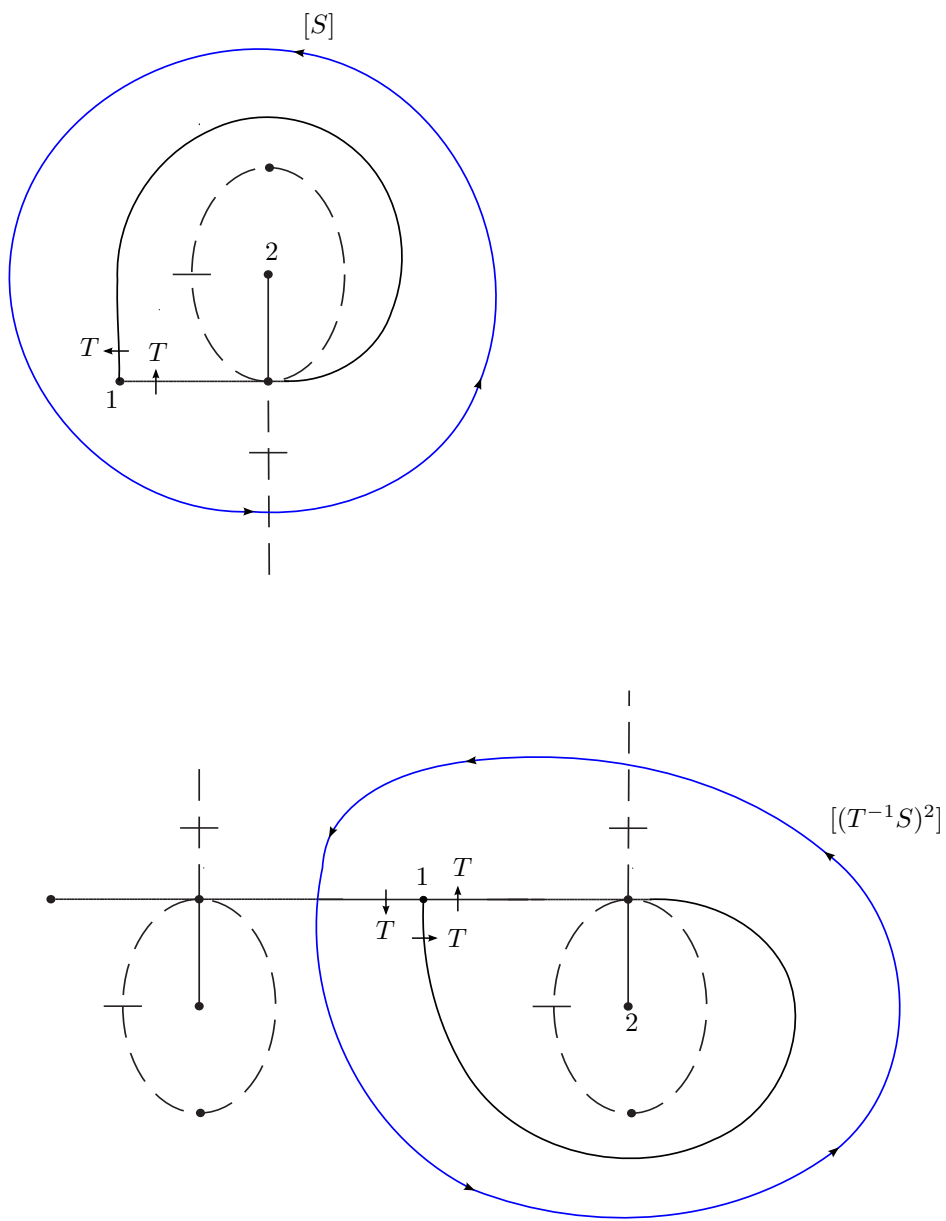}
\caption{No BPS strings exist between points labelled 1 and 2 for the cases $[S]$ and $[(T^{-1}S)^2]$ with $SU(2)$, respectively $SU(3)$ made manifest.}\label{fig:noBPSstringADsings}
\end{figure}

For the $[S]$ and $[(T^{-1}S)^2]$ cases we can draw BPS strings between any pair of 7-branes inside the $[S]$ and $[(T^{-1}S)^2]$ loops of Fig. \ref{fig:conjugacyclassloops}. The generators
that lie outside the Cartan subalgebra correspond to strings starting
and ending on different 7-branes. For $SU(n)$ groups the
number of generators outside the Cartan subalgebra is $n(n-1)$. Hence, for the $[S]$ and $[(T^{-1}S)^2]$ cases we need one and three BPS strings (taking into account that each string can have two orientations), respectively. The fact that we can draw strings between any pair of 7-branes means that there are more BPS strings than gauge group generators. This situation is resolved (and not just improved) by putting some of the 7-branes on top of each other. For the $[S]$ and $[(T^{-1}S)^2]$ cases we can put two and three 7-branes coincident, respectively, see Fig. \ref{fig:noBPSstringADsings}. Now we have manifest $SU(2)$ and $SU(3)$ symmetries and it can be shown that just as in the $[T^{-1}S]$ case there do not exist any BPS strings that lie stretched between the stack of coinciding 7-branes and the 7-brane that is locked by the $\pm S$ branch cuts nor do there exist strings that go from the stack back to itself around some non-contractible loop.

\subsection{$D$ series}\label{Dseries}

\begin{figure}
\centering
\psfrag{T1}{$T$}
\psfrag{T2}{$T$}
\psfrag{T3}{$T$}
\psfrag{T4}{$T$}
\psfrag{[S]}{$[S]$}
\psfrag{[-1]}{$[-\mathbbm{1}]$}
\psfrag{1}{$1$}
\psfrag{2}{$2$}
\psfrag{3}{$3$}
\includegraphics[scale=.8]{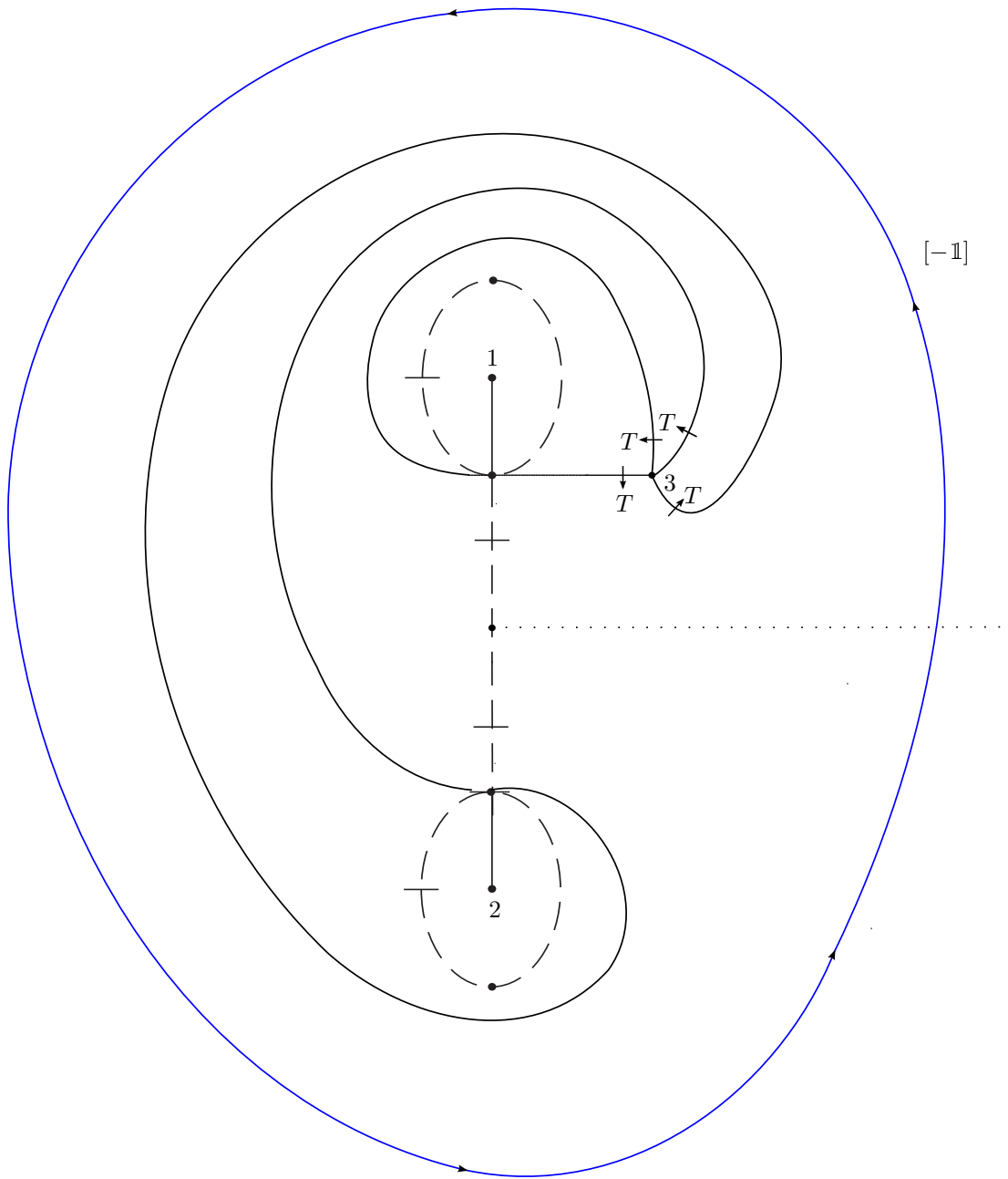}
\caption{Branch cuts for a manifest $SU(4)$ inside the $[-\mathbbm{1}]$ loop.}\label{fig:branchcutssu(4)}
\end{figure}

\begin{figure}
\centering
\psfrag{1}{$1$}
\psfrag{2}{$2$}
\psfrag{3}{$3$}
\includegraphics{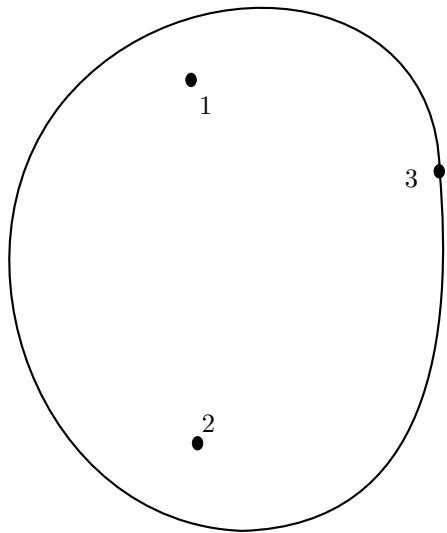}
\caption{Shown is the only BPS string that still exists in the $SO(8)$ case when an $SU(4)$ symmetry group is made manifest. This BPS string goes from the stack of four 7-branes at the point labelled 3 back to itself encircling the single branes 1 and 2.}\label{fig:BPSSO(8)}
\end{figure}

The relation between BPS open strings and gauge group generators for the $D_4$ case can be studied by making an $SU(4)$ subgroup manifest. Fig. \ref{fig:branchcutssu(4)} shows the branch cut representation when four 7-branes are made to coincide inside the $[-\mathbbm{1}]$ loop encircling six 7-branes. The only BPS string that still exists in this case is the one drawn in Fig. \ref{fig:BPSSO(8)}.

Fig. \ref{fig:BPSSO(8)} represents the three locations of the 7-branes of Fig. \ref{fig:branchcutssu(4)} but without the branch cuts. Strictly speaking one should also draw those but in order not to make the pictures too messy we leave them out. Branes 1 and 2 correspond to the locked 7-branes of Fig. \ref{fig:branchcutssu(4)}. Brane 3 corresponds to the stack of four 7-branes. It turns out that the only admissible BPS string is the one that starts and ends on the stack of four 7-branes and that loops around the other two locked 7-branes.

The generators outside the $SU(4)$ subgroup of $SO(8)$ can be represented by their $SU(4)$ representation. We have the following branching rule for the decomposition of the adjoint representation of $SO(8)$ in terms of irreps of the subalgebra $SU(4)\times U(1)$:
\begin{equation}
 \mathbf{28} \rightarrow \mathbf{1}+\mathbf{15}+\mathbf{6}+\bar{\mathbf{6}}\,,
\end{equation}
where we have suppressed the $U(1)$ labels because we will not be able to relate those to properties of the BPS strings anyway. The $SU(4)$ singlet $\mathbf{1}$ is one of the elements of the Cartan subalgebra of $SO(8)$. The $\mathbf{15}$ of $SU(4)$ is made manifest and does not require any massive BPS strings. We are left with the $\mathbf{6}$ and $\bar{\mathbf{6}}$ of $SU(4)$ which are the antisymmetric rank two tensor of $SU(4)$ and its conjugate, respectively.

The BPS string drawn in Fig. \ref{fig:BPSSO(8)} has $\Lambda=-\mathbbm{1}$. This means that the 7-brane on which the string has started and on which it can end must be different branes in the stack. Counting both orientations along the string there are 4 times 3, i.e.~12 such strings. For each orientation there are thus 6 such strings. These are the sought for $\mathbf{6}$ and $\bar{\mathbf{6}}$ irreps of $SU(4)$. We see that conjugation of the $SU(4)$ irrep corresponds to orientation reversal along the string.

Instead of making $SU(4)$ manifest we could also have made, say, an $SU(2)\times SU(2)\times SU(2)$ subgroup of $SO(8)$ manifest using a different branch cut representation. The reason that $SU(4)$ is attractive is because the generators outside the $U(4)$ subgroup of $SO(8)$ are antisymmetric rank two tensors and for such irreps we know how to relate them (in contrast to for example singlet representations) to the BPS strings. In perturbative string theory the $U(4)$ subgroup is clearly the natural one to explain the emergence of $SO(8)$. Also in the split orientifold case studied in \cite{Sen:1996vd} the $SU(4)$ group plays an important role. It is not a priori guaranteed that there exists an open string description of $SO(8)$ via some other subgroup such as $SU(2)\times SU(2)\times SU(2)$. It can for example happen that making such a symmetry manifest leads to fewer BPS strings than generators. When this is the case such a construction will not work. This is in fact what happens when $SU(2)\times SU(2)\times SU(2)$ is made manifest.

The $D_{n+4}$-type gauge groups are related to the $[-T^n]$ conjugacy classes. The set of 7-branes giving rise to $SO(10)$ are encircled in Fig. \ref{fig:conjugacyclassloops} by the $[-T]$ loop. In this case in order to have a one-to-one relation between BPS strings and gauge group generators we have to make an $SU(5)$ subgroup manifest. The analysis proceeds analogously to the $D_4$ case and will not be given here. For the $D_{n+4}$ case we must make an $SU(n+4)$ subgroup manifest.

\subsection{$E$ series}\label{Eseries}

In the case of the $D_{n+4}$ series making $SU(n+4)$ manifest we always have one antisymmetric rank two tensor irrep of $SU(n+4)$ and one BPS string going from the stack of $n+4$ branes back to itself. As we will see for the case of the $E_n$ series with $n=6,7,8$ making for example a certain $SU(m)\times (U(1))^k$ subgroup with $n=m-1+k$ manifest will in general lead to more than one antisymmetric rank two tensor irrep of the $SU(m)$ subgroup. These antisymmetric rank two tensors differ only in their $k$ $U(1)$ charges of the $U(1)$'s in the decomposition $E_n\rightarrow SU(m)\times (U(1))^k$. Since these $U(1)$ charges cannot be made manifest one may wonder how we can reliably state that the number of inequivalent BPS strings going from the stack to itself equals the number of antisymmetric rank two tensor irreps of $SU(m)$. The reason is the following.

Suppose we have two strings $a$ and $b$ lying along the paths $\gamma_a$ and $\gamma_b$, respectively, each of which is non-contractible starting and ending on the stack. Suppose that the BPS strings $a$ and $b$ have exactly the same Chan--Paton labels with respect to the $SU(m)\times (U(1))^k$ subgroup. Then it follows that the string lying along $\gamma_b^{-1}\gamma_a$ has trivial Chan--Paton labels and must correspond to a Cartan generator. These are formed by massless strings starting and ending on the stack. Inside the region, $D$, bounded by the loops encircling the collapsable 7-brane configurations of Table \ref{Tatealgorithm} such massless strings lie along contractible paths (in the sense of the homotopy of $D$) and hence if strings $a$ and $b$ have the same Chan--Paton labels then $\gamma_b^{-1}\gamma_a$ must be contractible or in other words $\gamma_a$ and $\gamma_b$ are homotopically equivalent. Therefore, homotopically inequivalent strings starting and ending on the same stack must have different sets of $k$ $U(1)$ charges and it follows that the number of such inequivalent BPS strings must match the number of antisymmetric rank two tensors in the decomposition of some $E_n$ group according to $E_n\rightarrow SU(m)\times (U(1))^k$. We will verify that this is indeed the case.

The matching of stack-to-stack-strings with generators transforming as antisymmetric rank two tensors of $SU(m)$ provides a nontrivial step towards an open string interpretation of the exceptional gauge groups. However, since in the decomposition of the adjoint representation of $E_n$ according to $E_n\rightarrow SU(m)\times (U(1))^k$ there also appear states that are in the fundamental or singlet of $SU(m)$ it is difficult to match all the generators to strings because for strings with only one or no endpoints on the stack we have no general argument to relate them in number to the fundamental or singlet irreps of $SU(m)$. Still, for each of the cases $E_n$ with $n=6,7,8$, as we will see below, it is possible to perform certain consistency checks regarding the number of such strings.

We will next motivate why we make $SU(m)\times (U(1))^k$ subgroups of $E_n$ manifest and fix what $m$ and $k$ should be for each of the cases $n=6,7,8$. The higher the symmetry we make manifest the less BPS strings there generically will be. In general when we make $SU(k)\times SU(l)$ subgroups manifest the adjoint of $E_n$ decomposes into irreps that are e.g.~in the fundamental of $SU(k)$ and in the antisymmetric rank two of $SU(l)$. Such states cannot be realized using open strings. If we make $SU(m)$ subgroups manifest with a relatively high value of $m$, such as $m=6$ for $E_6$ or $m=7$ for $E_7$ then there appear typically higher rank than two antisymmetric tensor representations and also these cannot be realized with open strings. For branching rules of the exceptional symmetry groups we refer to e.g.~\cite{Slansky:1981yr}. Open strings can only account for singlets, fundamental and antisymmetric rank two tensor irreps and only in such a way that their conjugates also appear. This is because for each BPS string we always have both orientations.

In general the branching rule of the adjoint decomposition of some to be formed gauge group with respect to some $SU$ subgroup (or possibly direct product of $SU$ subgroups) depends on the embedding. A simple condition the subgroup must satisfy is that its rank must equal that of the gauge group. Therefore the embedding always goes via a maximal subalgebra. We will choose subgroups of the $E$-type gauge groups for which the adjoint decomposition always gives the same irreps of the subgroup (ignoring possible differences in $U(1)$ charges) regardless via which maximal subalgebra it is embedded. Having branching rules independent of the embedding is convenient because it means that when we study BPS strings we do not need to worry about the question which embedding is realized by a certain branch cut representation compatible with some manifest $SU$ group.

In order that the adjoint decomposition of $E_n$ only leads to open string realizable states such that the irreps (apart from $U(1)$ charges) are the same regardless the embedding (that must always go via a maximal subalgebra) we choose the $SU(m)\times (U(1))^k$ subgroups of $E_n$ of Table \ref{Esubgroups}. We will thus work with a manifest $SU(5)$ symmetry group.
\begin{table}[h!]
\begin{center}
\begin{tabular}{ccc}
  \hline
 $n$  & $m$  & $k$   \\
  \hline
6 & 5 & 2 \\
7 & 5 & 3 \\
8 & 5 & 4\\
  \hline
\end{tabular}
\end{center}
\caption{{\small{$SU(m)\times (U(1))^k$ subgroups of $E_n$ for which the adjoint decomposition of $E_n$ gives the same number of open string realizable irreps of $SU(m)$ for every embedding of $SU(m)\times (U(1))^k$ into $E_n$ with $n=m-1+k$.}}}
\label{Esubgroups}
\end{table}

Any BPS string that exists inside a $[-T^{-1}S]$, $[-S]$ or $[-(T^{-1}S)^2]$ loop with a manifest $SU(5)$ automatically carries a representation with respect to $SU(5)$ and these representations appear in the adjoint decomposition of $E_n$. Granted there exists an open string description for the exceptional singularities we have the following consistency conditions for these open strings.
\begin{enumerate}
\renewcommand{\labelenumi}{\Roman{enumi}.}
\item For the strings that start and end on the stack with $\Lambda=-\mathbbm{1}$, by the argument given at the beginning of this section, we know that, regardless the branch cut representation, there should always be as many such strings as there are antisymmetric rank two tensors of $SU(5)$ in the adjoint decomposition of $E_n$.
\newcounter{last}
\setcounter{last}{\value{enumi}}
\end{enumerate}
For the strings with at least one endpoint on a single brane we cannot state for a given branch cut structure exactly how many such strings there should be. In fact the number of such strings varies depending on the positioning of the branch cuts\footnote{The number of branes on the same branch is the number of branes that can be connected by non-selfintersecting $\Lambda=+\mathbbm{1}$ paths that do not cross any branch cuts. This number determines which $SU$ symmetry groups corresponding to the $[T^n]$ conjugacy classes can be realized without changing the positioning of the $S$ branch cuts. BPS strings that cross the direct paths between such branes no longer exist once these branes are put on top of each other. Since making a certain $SU$ symmetry group manifest can in general be done in different ways depending on the positioning of the $S$ branch cuts the number of BPS strings can differ for different branch cut representations each of which makes, with the same value for $m$, an $SU(m)$ group manifest.}. A simple consistency condition is then
\begin{enumerate}
\setcounter{enumi}{\value{last}}
\renewcommand{\labelenumi}{\Roman{enumi}.}
\item There are always at least as many stack-to-single-brane-strings as there are fundamentals in the adjoint decomposition of $E_n$ and there are at least as many single-to-single-brane-strings (counting both orientations) as there are singlets outside the Cartan subalgebra.
\end{enumerate}
\bigskip

\noindent We will now discuss the cases of $E_6, E_7$ and $E_8$ separately.
\bigskip

\noindent $\bullet\ \  E_6$
\medskip

\begin{figure}
\centering
\psfrag{[minS]}{$[-S]$}
\psfrag{[minTinvS]}{$[-T^{-1}S]$}
\psfrag{[minTinvSsquare]}{$[-(T^{-1}S)^2]$}
\psfrag{T1}{$T$}
\psfrag{T2}{$T$}
\psfrag{T3}{$T$}
\psfrag{T4}{$T$}
\psfrag{T5}{$T$}
\psfrag{1}{$1$}
\psfrag{2}{$2$}
\psfrag{3}{$3$}
\psfrag{4}{$4$}
\psfrag{5}{$5$}
\psfrag{6}{$6$}
\includegraphics[scale=.8]{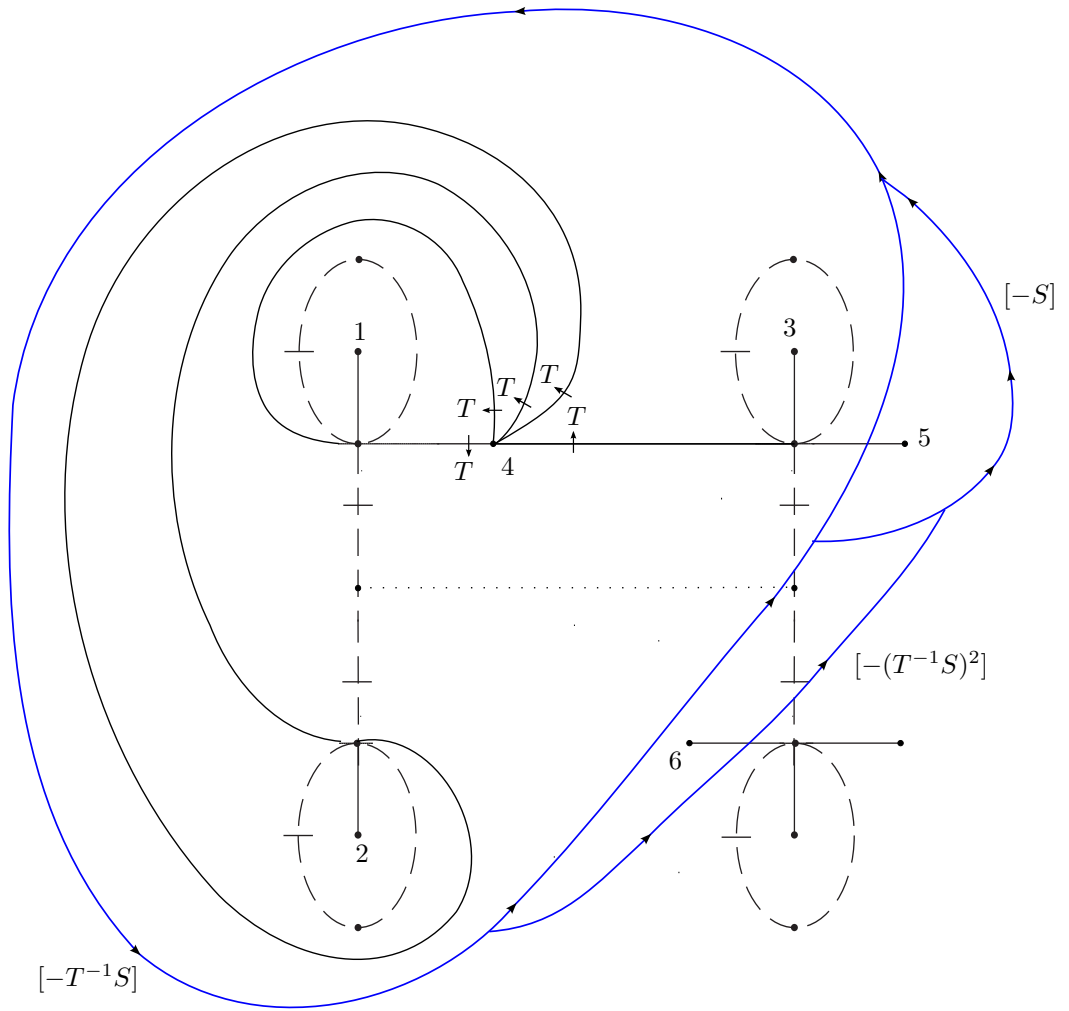}
\caption{Branch cuts for a manifest $SU(5)$ inside the $[-T^{-1}S]$, $[-S]$ and $[-(T^{-1}S)^2]$ loops.}\label{fig:branchcutsEgroups}
\end{figure}

We start with the case of the $E_6$ gauge group. When we make an $SU(5)$ symmetry manifest one possible branch cut representation is the one given in Fig.~\ref{fig:branchcutsEgroups} where for $E_6$ we should consider the $[-T^{-1}S]$ loop. In Fig.~\ref{fig:branchcutsEgroups} there are three locked 7-branes denoted by 1, 2 and 3 and there is a stack of five 7-branes located at the point labelled 4. Because there is now a manifest $SU(5)$ group, 20 of the 72 generators outside the Cartan subalgebra of $E_6$ are taken care off.

The decomposition of the adjoint representation of $E_6$ according to the subalgebra $SU(5)\times (U(1))^2$ is
 \begin{equation}
  \mathbf{78}\rightarrow\mathbf{24}+2\cdot\mathbf{1}+2\cdot\left(\mathbf{10}+\bar{\mathbf{10}}\right)+\mathbf{5}+
\bar{\mathbf{5}}+2\cdot\mathbf{1}\,.
 \end{equation}
The $U(1)$ charges have been suppressed. By $2\cdot\mathbf{1}$ we mean two singlets and by $2\cdot\left(\mathbf{10}+\bar{\mathbf{10}}\right)$ we mean two sets of $\mathbf{10}+\bar{\mathbf{10}}$ irreps. There are in total four singlets separated in two sets of two. The $U(1)$ charges can still depend on the embedding. We only need two facts about these charges that are independent of the embedding. The first set of states $2\cdot\mathbf{1}$ contains two singlets whose $U(1)$ charges are all equal to zero and corresponds to two Cartan generators while the second set of states $2\cdot\mathbf{1}$ contains two singlets that have nonzero $U(1)$ charges, with the $U(1)$ charges of one singlet opposite to those of the other singlet, and corresponds to generators outside the Cartan subalgebra.

\begin{figure}
\centering
\psfrag{1}{$1$}
\psfrag{2}{$2$}
\psfrag{3}{$3$}
\psfrag{4}{$4$}
\psfrag{a}{$a$}
\psfrag{b}{$b$}
\includegraphics[scale=.8]{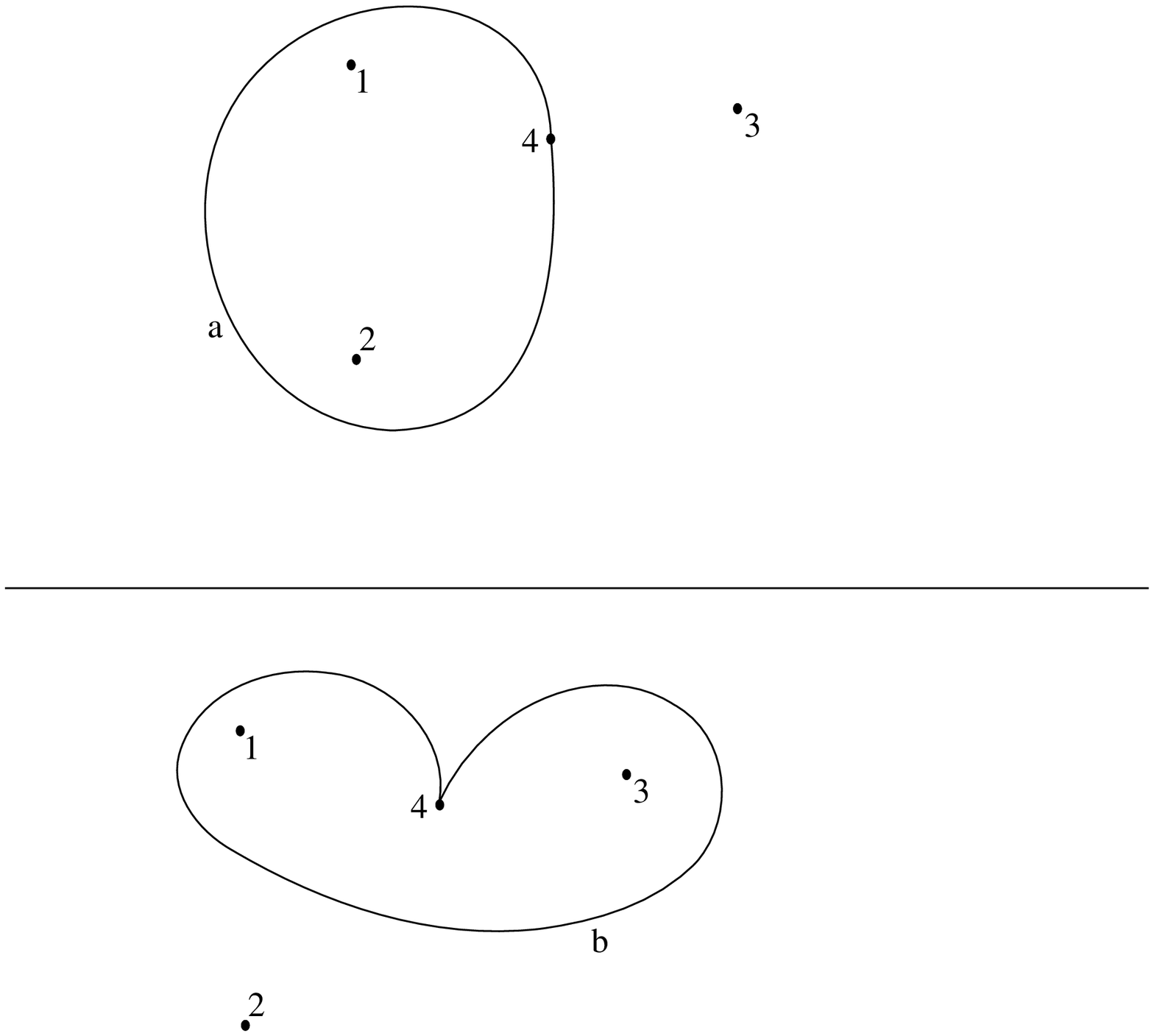}
\caption{All the stack-to-stack-strings with $\Lambda=-\mathbbm{1}$ for $E_6$.}\label{fig:stackstackE6}
\end{figure}

From the argument at the beginning of this subsection we know that the number of antisymmetric rank two tensors equals the number of inequivalent strings that start and end on the stack of five 7-branes. By inspection it can be seen that there are two such strings, see Fig.~\ref{fig:stackstackE6}, namely:
\begin{enumerate}
\renewcommand{\labelenumi}{\alph{enumi}.}
\item From 4 to 4 with $\Lambda=-\mathbbm{1}$ around 1 and 2.
\item From 4 to 4 with $\Lambda=-\mathbbm{1}$ around 1 and 3.
\end{enumerate}
The two BPS strings going from 4 to 4 have $\Lambda=-\mathbbm{1}$ so that they cannot start and end on the same brane in the stack. Therefore, counting both orientations, each of these BPS strings gives rise to 20 generators. Since orientation reversal corresponds to conjugation each BPS string corresponds to one set of $\mathbf{10}+\bar{\mathbf{10}}$ generators. BPS strings a and b are homotopically distinct and must therefore carry different $U(1)$ charges.

\begin{figure}
\centering
\vskip -2cm
\psfrag{a}{$1$}
\psfrag{b}{$2$}
\psfrag{c}{$3$}
\psfrag{d}{$4$}
\psfrag{1}{$1$}
\psfrag{2}{$2$}
\psfrag{3}{$3$}
\psfrag{4}{$4$}
\psfrag{I}{$1$}
\psfrag{II}{$2$}
\psfrag{III}{$3$}
\psfrag{IV}{$4$}
\includegraphics[scale=.8]{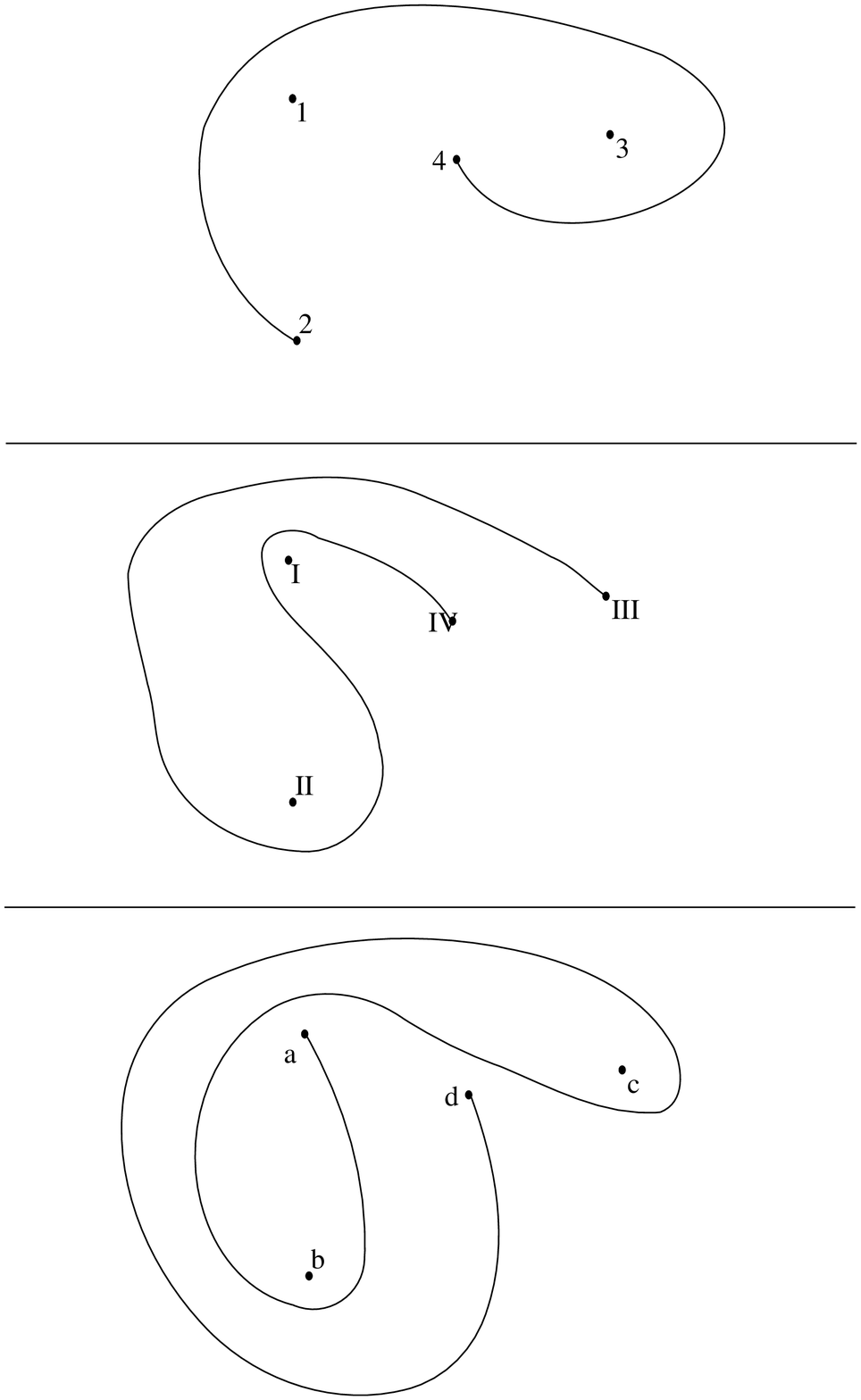}
\caption{Stack-to-single-brane-strings with $\Lambda=\pm\mathbbm{1}$ for $E_6$.}\label{fig:stacksingleE6}
\end{figure}

\begin{figure}
\centering
\psfrag{1}{$1$}
\psfrag{2}{$2$}
\psfrag{3}{$3$}
\psfrag{4}{$4$}
\includegraphics[scale=.8]{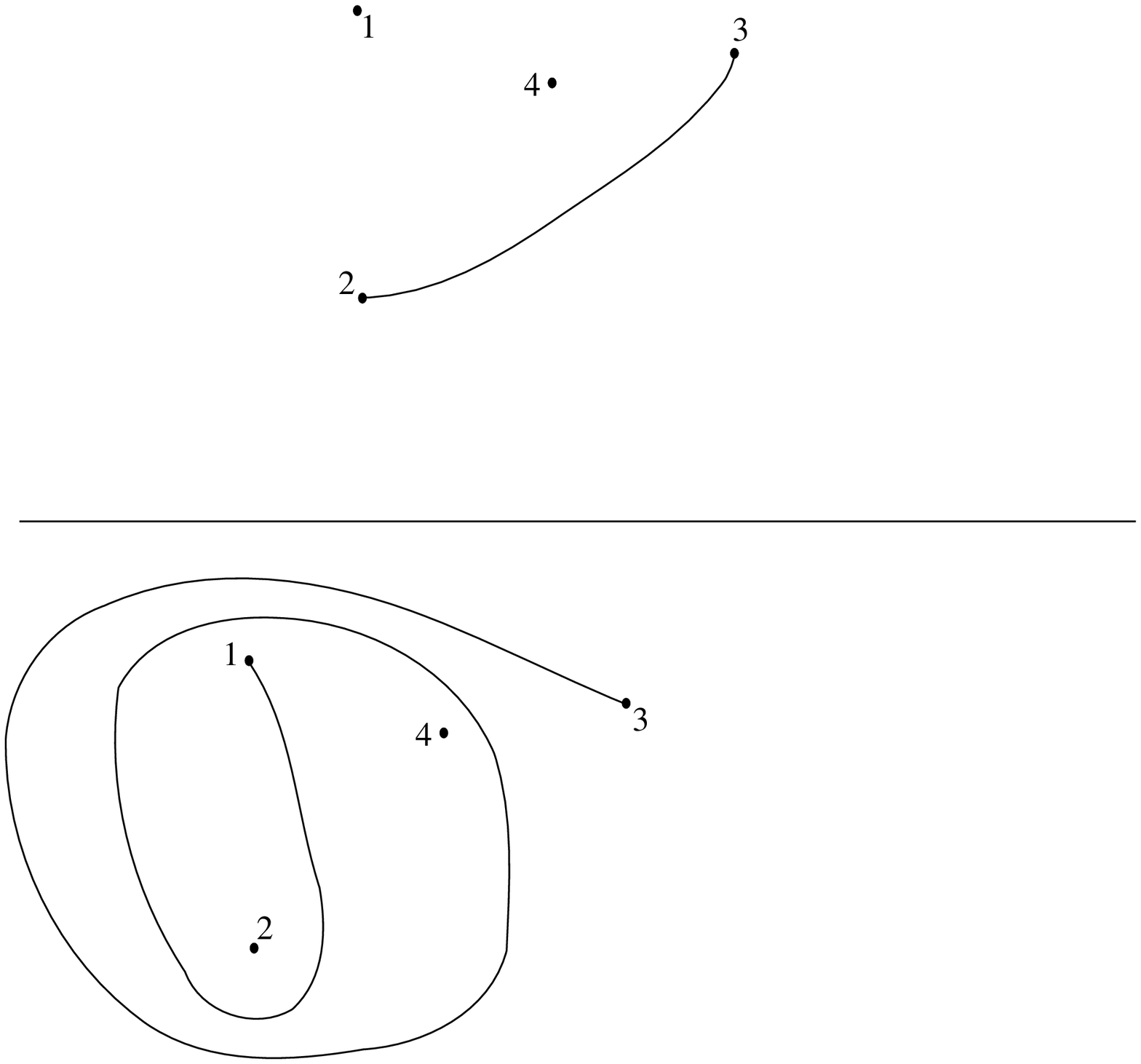}
\caption{Single-to-single-brane-strings with $\Lambda=\pm\mathbbm{1}$ for $E_6$.}\label{fig:singlesingleE6}
\end{figure}

This leaves us with the challenge of relating the $\mathbf{5}+\bar{\mathbf{5}}+2\cdot\mathbf{1}$ states to BPS strings. What we can say with certainty is that  the $\mathbf{5}+\bar{\mathbf{5}}$ states correspond to strings with one endpoint on the stack of five 7-branes and one endpoint on a single brane and that the $2\cdot\mathbf{1}$ states correspond to strings that lie stretched between two different single branes. By inspection it can be checked that such BPS strings exist in sufficient number. As shown in Fig.~\ref{fig:stacksingleE6} there are BPS strings starting at 4 and going (along suitable paths) to any of the points 1, 2 or 3. Fig.~\ref{fig:singlesingleE6} shows that there also exist strings going from brane 3 to 2 and from 3 to 1. Hence, the required type of strings can be constructed but we do not know how to relate them to the gauge group generators.  We verified that the branch cut representation with manifest $SU(5)$ symmetry of 
Fig.~\ref{fig:branchcutsEgroups} allows for BPS strings that satisfy the above-mentioned consistency conditions I and II.

\begin{figure}
\centering
\psfrag{T}{$T$}
\psfrag{[minS]}{$[-S]$}
\psfrag{[minTinvS]}{$[-T^{-1}S]$}
\includegraphics[scale=.8]{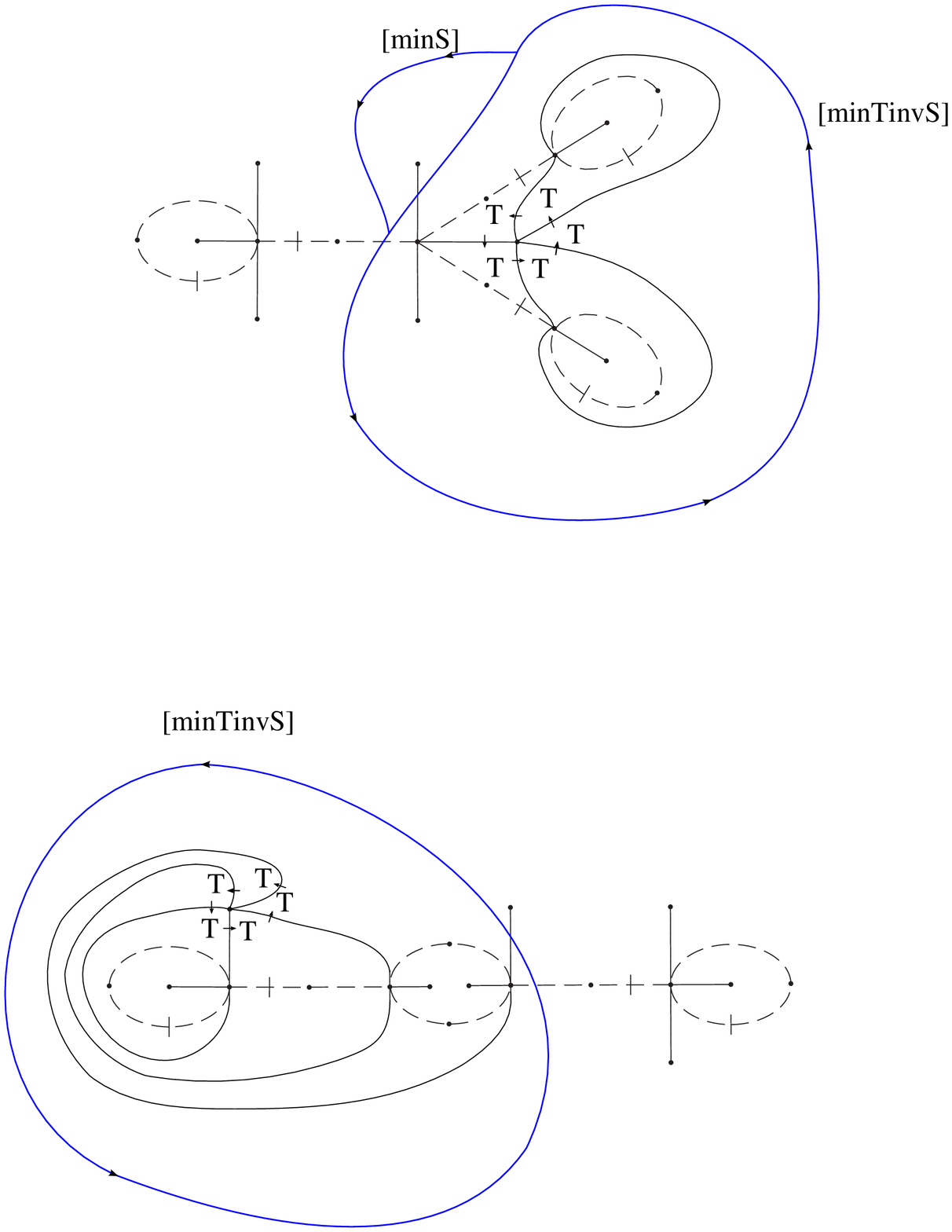}
\caption{Two alternative ways of making $SU(5)$ manifest inside loops relevant for the exceptional symmetry groups.}\label{fig:alternatives}
\end{figure}

To see the effect of choosing different branch cuts consider Fig.~\ref{fig:alternatives} which shows two alternative ways of placing the $S$ branch cuts while having an $SU(5)$ inside a $[-T^{-1}S]$ loop. In both cases it can be verified that conditions I and II are met. In both cases there are two homotopically inequivalent $\Lambda=-\mathbbm{1}$ loops from the stack to itself and there are sufficiently many stack-to-single-brane- as well as single-to-single-brane-strings to in principle account for all the states in the adjoint decomposition of $E_6$.

The two branch cut representations of Fig.~\ref{fig:alternatives} allow us to make an $SU(5)\times SU(2)$ and an $SU(6)$ symmetry manifest. In the lower figure of Fig.~\ref{fig:alternatives} we can make an $SU(2)$ manifest by putting the two 7-branes that are encircled by the same $S$ branch cuts on top of each other. In the upper figure of Fig.~\ref{fig:alternatives} there exists a path from the stack of five branes to a single brane that does not cross any branch cuts. We can thus take the single brane along this path and put it on top of the stack forming a stack of six branes (without changing the $S$ branch cuts). In the decomposition of the adjoint representation of $E_6$ with respect to either $SU(5)\times SU(2)\times U(1)$ or $SU(6)\times U(1)$, which are given by ($U(1)$ charges suppressed)
\begin{eqnarray}
 \mathbf{78} & \rightarrow & (\mathbf{24},\mathbf{1})+(\mathbf{1},\mathbf{3})+(\mathbf{1},\mathbf{1})+(\mathbf{10},\mathbf{2})+(\bar{\mathbf{10}},\mathbf{2})+(\mathbf{5},\mathbf{1})+(\bar{\mathbf{5}},\mathbf{1})\,,\\
\mathbf{78} & \rightarrow & \mathbf{35}+\mathbf{1}+2\cdot\mathbf{20}+2\cdot\mathbf{1}\,,
\end{eqnarray}
respectively, there occur non-open string realizable representations of the manifest symmetry groups such as the  $(\mathbf{10},\mathbf{2})$ of $SU(5)\times SU(2)$ and the $\mathbf{20}$ of $SU(6)$. Hence when $SU(5)\times SU(2)$ or $SU(6)$ are made manifest we should not expect there to be an open string interpretation of the symmetry enhancement. Indeed when $SU(5)\times SU(2)$ is manifest we cannot construct a $\mathbf{5}+\bar{\mathbf{5}}$ string that necessarily would have to go from the stack of five 7-branes to the only single brane. Likewise when $SU(6)$ is manifest we cannot construct strings corresponding to singlets which necessarily would have to lie between the only two single branes.

One could try to use multi-pronged strings to describe representations such as $(\mathbf{10},\mathbf{2})+(\bar{\mathbf{10}},\mathbf{2})$. A strategy could be to take the two branes forming $SU(2)$ apart so that now stack-to-stack-strings in the
$\mathbf{10}+\bar{\mathbf{10}}$ of $SU(5)$ exist and to transform those into multi-pronged strings using rules similar to those used in \cite{Gaberdiel:1997ud,Gaberdiel:1998mv} and then to put the branes on top of each other after the multi-pronged strings have been formed. It would be interesting to work out the details of such an analysis using our global branch cut structure.

We pause here to make a comment on statements regarding (non-)existence of certain BPS strings that are made at various places in the text. When we say that there are no strings of a certain type or no more than drawn in one of the figures this means that we did not manage to construct those after an extensive search. We did not try to prove these statements in a rigorous way. Such an attempt would probably greatly benefit from some computer program that computes $SL(2,\mathbb{Z})$ transformations going from one connected region of the domain of $(\tau,f)$ to an adjacent one tracing out paths such that no self-intersections occur.
\medskip

\noindent $\bullet\ \ E_7$
\medskip

For $E_7$ the adjoint decomposes into irreps of $SU(5)\times (U(1))^3$ as follows:
 \begin{equation}
 \mathbf{133}\rightarrow\mathbf{24}+3\cdot\mathbf{1}+3\cdot\left(\mathbf{10}+\bar{\mathbf{10}}\right)+4\cdot\left(\mathbf{5}+\bar{\mathbf{5}}\right)+6\cdot\mathbf{1}\,,
 \end{equation}
where we have suppressed the $U(1)$ labels. The first set of singlets corresponds to Cartan generators while the second set of singlets to generators outside the Cartan subalgebra. This second set of singlets, denoted by $6\cdot\mathbf{1}$, can be divided into two sets where one set has opposite $U(1)$ charges with respect to the other set.

Independent of the branch cut representation according to rules I and II we need exactly three homotopically inequivalent strings from the stack to itself, four or more strings going from the stack to a single brane and three or more strings stretched between different single branes.

\begin{figure}
\centering
\psfrag{1}{$1$}
\psfrag{2}{$2$}
\psfrag{3}{$3$}
\psfrag{4}{$4$}
\psfrag{5}{$5$}
\psfrag{c}{$c$}
\includegraphics{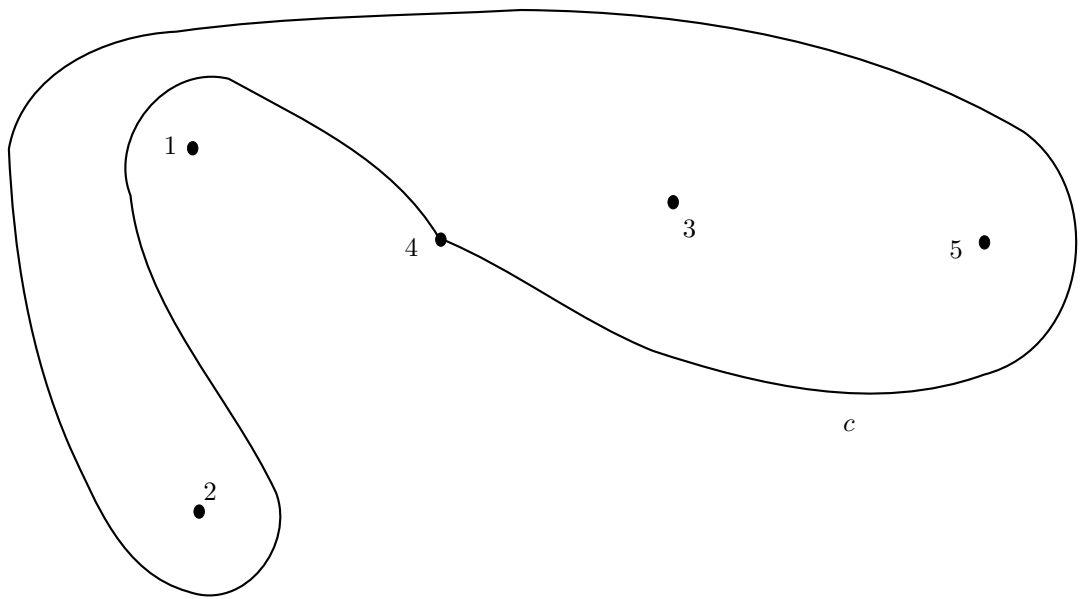}
\caption{All the stack-to-stack-strings with $\Lambda=-\mathbbm{1}$ that exist for $E_7$ but not for $E_6$.}\label{fig:stackstackE7}
\end{figure}

Consider the branch cut representation for a manifest $SU(5)$ of Fig.~\ref{fig:branchcutsEgroups} where we take the $[-S]$ loop.
The antisymmetric rank two irreps must be related in number to strings with $\Lambda=-\mathbbm{1}$ that start and end on the stack. These BPS strings are:
\begin{enumerate}
\renewcommand{\labelenumi}{\alph{enumi}.}
\item From 4 to 4 with $\Lambda=-\mathbbm{1}$ around 1 and 2.
\item From 4 to 4 with $\Lambda=-\mathbbm{1}$ around 1 and 3.
\item From 4 to 4 with $\Lambda=-\mathbbm{1}$ around 2, 3 and 5.
\end{enumerate}
Each of the strings a to c, counting both orientations, accounts for one set of $\mathbf{10}+\bar{\mathbf{10}}$ generators. String a also exists in the $D_4$ case, and strings a and b exist in the $E_6$ case. Fig.~\ref{fig:stackstackE7} shows string c that does not exist for the $E_6$ case.

\begin{figure}
\centering
\psfrag{1}{$1$}
\psfrag{2}{$2$}
\psfrag{3}{$3$}
\psfrag{4}{$4$}
\psfrag{5}{$5$}
\includegraphics[scale=.7]{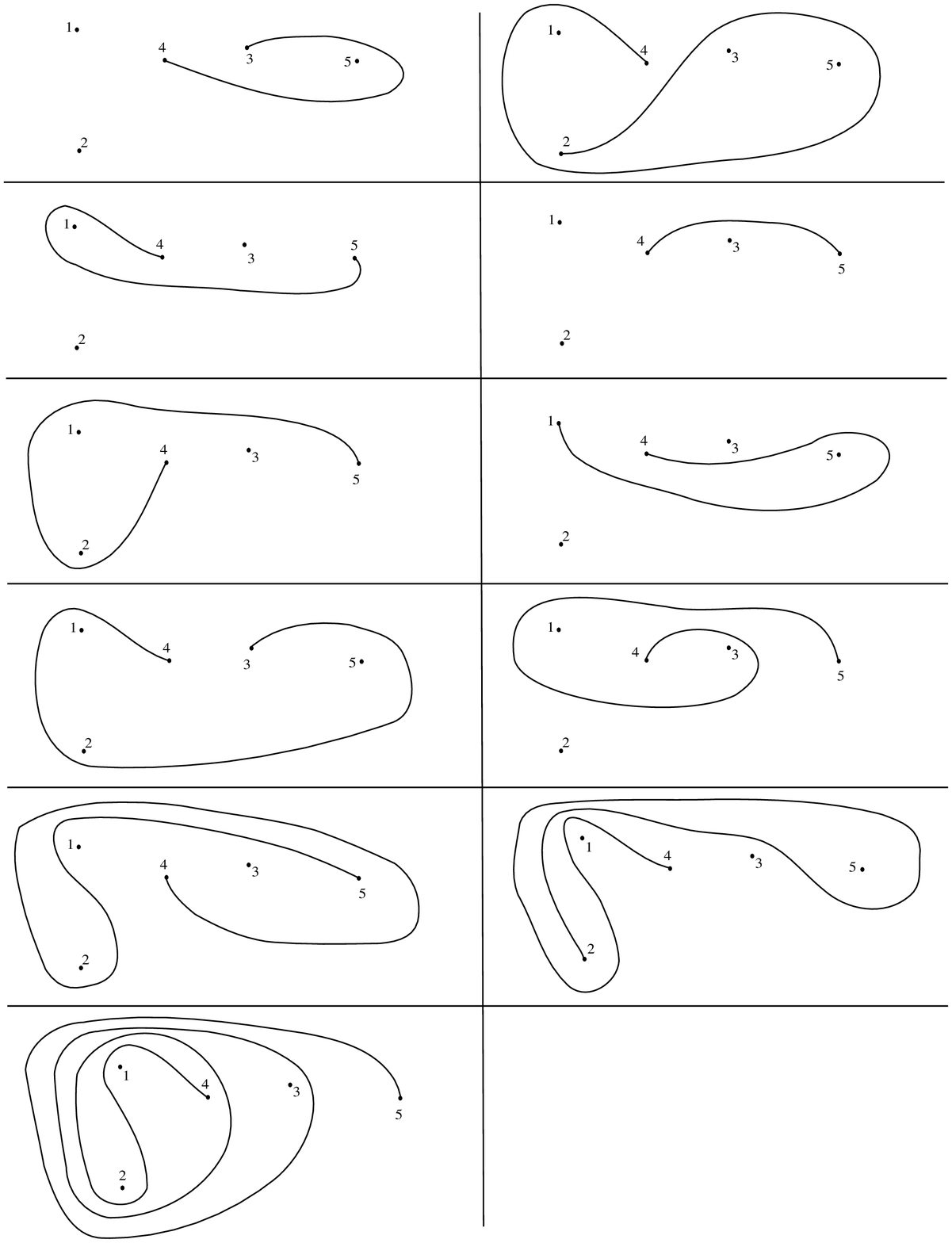}
\caption{Stack-to-single-brane-strings with $\Lambda=\pm\mathbbm{1}$ that exist for $E_7$ but not for $E_6$.}\label{fig:stacksingleE7}
\end{figure}

\begin{figure}
\centering
\psfrag{1}{$1$}
\psfrag{2}{$2$}
\psfrag{3}{$3$}
\psfrag{4}{$4$}
\psfrag{5}{$5$}
\includegraphics[scale=.7]{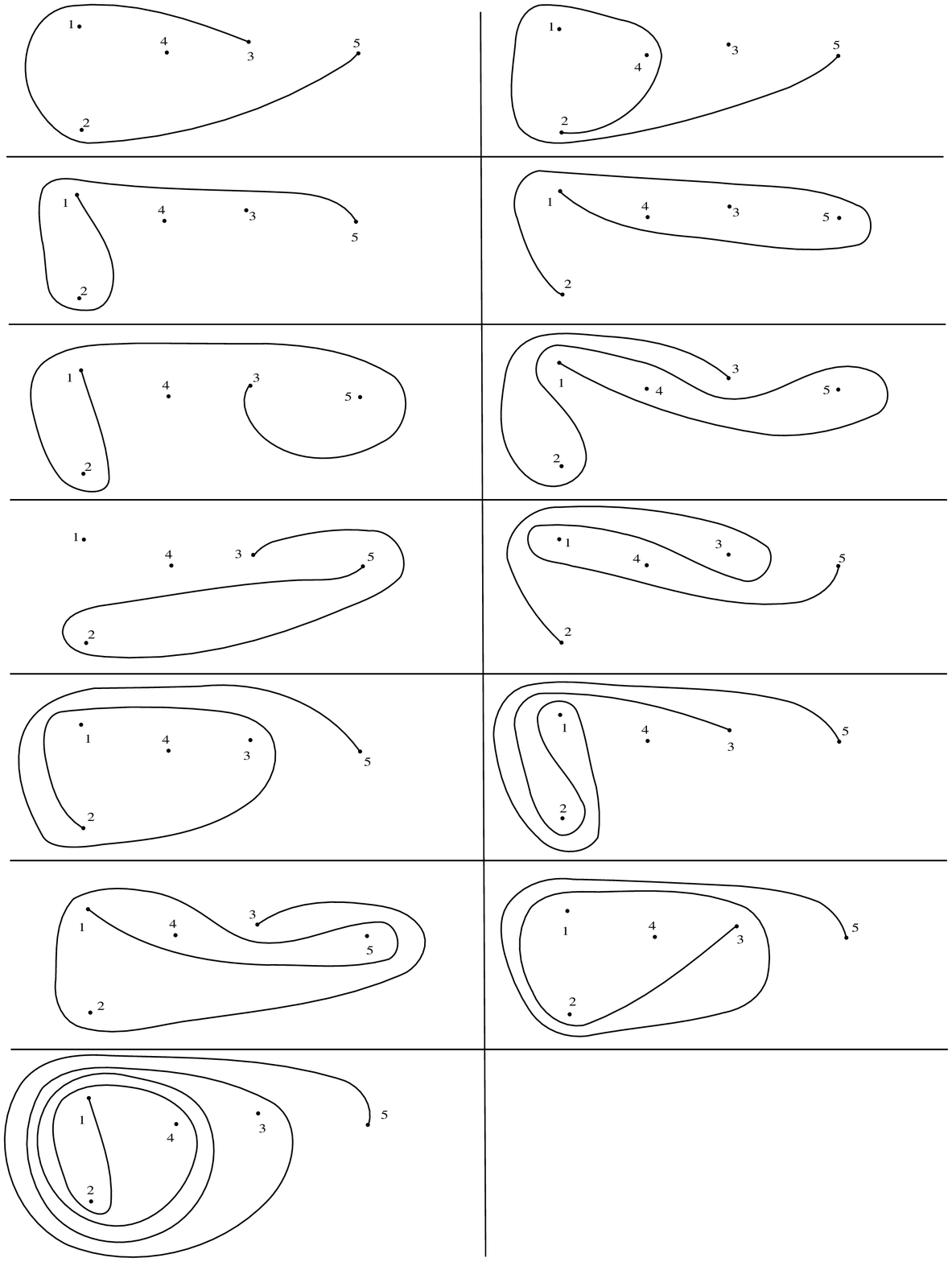}
\caption{Single-to-single-brane-strings with $\Lambda=\pm\mathbbm{1}$ that exist for $E_7$ but not for $E_6$.}\label{fig:singlesingleE7}
\end{figure}

The same is true for the stack-to-single-brane- and single-to-single-brane-strings that exist in the $E_6$ case. These also exist in the $E_7$ case. Stack-to-single-brane-strings that exist for $E_7$ but not for
$E_6$ are drawn in Fig.~\ref{fig:stacksingleE7} and single-to-single-brane-strings that exist for $E_7$ but not for $E_6$ are drawn in Fig.~\ref{fig:singlesingleE7}. Since these are sufficient in number and since there are exactly three homotopically inequivalent stack-to-stack-strings with $\Lambda=-\mathbbm{1}$ we once again verified conditions I and II.

We do not know if the strings drawn in Figs. \ref{fig:stacksingleE7} and \ref{fig:singlesingleE7} are really all the strings with at least one endpoint on a single brane. One way to check if there might be more BPS  strings is to combine certain strings with others to form new strings. The resulting path can be interpreted as a single new BPS string as long as it does not selfintersect and as long as there are no problems with charge conservation. For example a string going from a single brane back to itself along some non-contractible loop with $\Lambda=-\mathbbm{1}$ is not allowed. An example of an allowed combination is the joining of the first and third strings of Fig.~\ref{fig:singlesingleE7}. When this is done the resulting string is homotopically equivalent to the second string of Fig.~\ref{fig:singlesingleE6}. There are many examples of such combinations. We checked that the total set of $E_7$ strings given in
Figs. \ref{fig:stackstackE6} to \ref{fig:singlesingleE6} and Figs.
\ref{fig:stackstackE7} to \ref{fig:singlesingleE7} is closed under such combinations.

Besides the conditions I and II we can in the $E_7$ case perform another consistency check on the BPS strings. From the branch cut representation for $E_7$ shown in Fig.~\ref{fig:branchcutsEgroups} it is clear that we could make an $SU(6)$ symmetry manifest by putting brane 5 on top of the stack at 4. The path along which this is done is the profile of the fourth open string of Fig.~\ref{fig:stacksingleE7}. By looking at the stack-to-single-brane-strings of Figs.~\ref{fig:stacksingleE6} and \ref{fig:stacksingleE7} it can be concluded that all of them with the exception of the 3rd, 4th, 5th, 8th and 9th strings of Fig.~\ref{fig:stacksingleE7} disappear when $SU(6)$ is realized in this way. When brane 5 is put on top of the stack at 4 the 3rd and 8th strings of Fig.~\ref{fig:stacksingleE7} become identical. Each of the surviving strings of Fig.~\ref{fig:stacksingleE7} goes from the stack of five branes to brane number 5 and hence now go from the stack to itself but always end on the same brane in the stack. If we consider the stack-to-stack-strings of Figs.~\ref{fig:stackstackE6} and \ref{fig:stackstackE7} then first of all each of them survive putting brane 5 on top of the stack and secondly these strings do not end on brane 5. Combining them with those of Fig.~\ref{fig:stacksingleE7} we obtain all the stack-to-stack strings for a stack consisting of six branes and we thus find three sets of $\mathbf{15}+\bar{\mathbf{15}}$ irreps of $SU(6)$. Further, only three of the single-to-single-brane-strings of Figs. \ref{fig:singlesingleE6} and \ref{fig:singlesingleE7} survive. These are the first string of Fig.~\ref{fig:singlesingleE6} and the fourth and fifth strings of Fig.~\ref{fig:singlesingleE7}. There exists a decomposition of the adjoint of $E_7$ with respect to $SU(6)\times (U(1))^2$ that reads
\begin{equation}
\mathbf{133}\rightarrow\mathbf{35}+2\cdot\mathbf{1}+3\cdot\left(\mathbf{15}+\bar{\mathbf{15}}\right)+6\cdot\mathbf{1}\,.
\end{equation}
Each of these states are represented by the strings of Figs. \ref{fig:stackstackE6} to \ref{fig:singlesingleE6} and Figs. \ref{fig:stackstackE7} to \ref{fig:singlesingleE7} after brane 5 has been put on top of the stack. It can further be checked that the resulting branch cut representation does not allow for any stack-to-single-brane-strings consistent with the above decomposition.

The branching rule of the adjoint decomposition of $E_7$ with respect to $SU(6)\times (U(1))^2$ depends on the embedding. There exists an embedding of $SU(6)\times (U(1))^2$ that goes via $SU(8)$ that has a different branching rule. In this case the branching rule is
\begin{equation}
\mathbf{133}\rightarrow\mathbf{35}+2\cdot\mathbf{1}+\mathbf{15}+\bar{\mathbf{15}}+2\cdot\left(\mathbf{6}+\bar{\mathbf{6}}\right)+2\cdot\mathbf{20}+2\cdot\mathbf{1}\,.
\end{equation}
There now appears the non-open-string-realizable-state $\mathbf{20}$ of $SU(6)$. Clearly, this embedding is not described by Fig.~\ref{fig:branchcutsEgroups} with brane 5 on top of the stack. However, consider the right figure of Fig.~\ref{fig:alternatives} and the branes inside the $[-S]$ loop. There now exist more than two $\mathbf{6}+\bar{\mathbf{6}}$ stack-to-single-brane-strings as well as one stack-to-stack string with $\Lambda=-\mathbbm{1}$ representing the $\mathbf{15}+\bar{\mathbf{15}}$ irreps of $SU(6)$. This example makes explicit that the embedding of a symmetry group is related to the branch cut representation. Even though in this case there still exist some open strings these are not capable of describing the enhancement to $E_7$ starting from this embedding of $SU(6)$.
\medskip

\noindent $\bullet\ \ E_8$
\medskip

\begin{figure}
\centering
\psfrag{1}{$1$}
\psfrag{2}{$2$}
\psfrag{3}{$3$}
\psfrag{4}{$4$}
\psfrag{5}{$5$}
\psfrag{6}{$6$}
\includegraphics{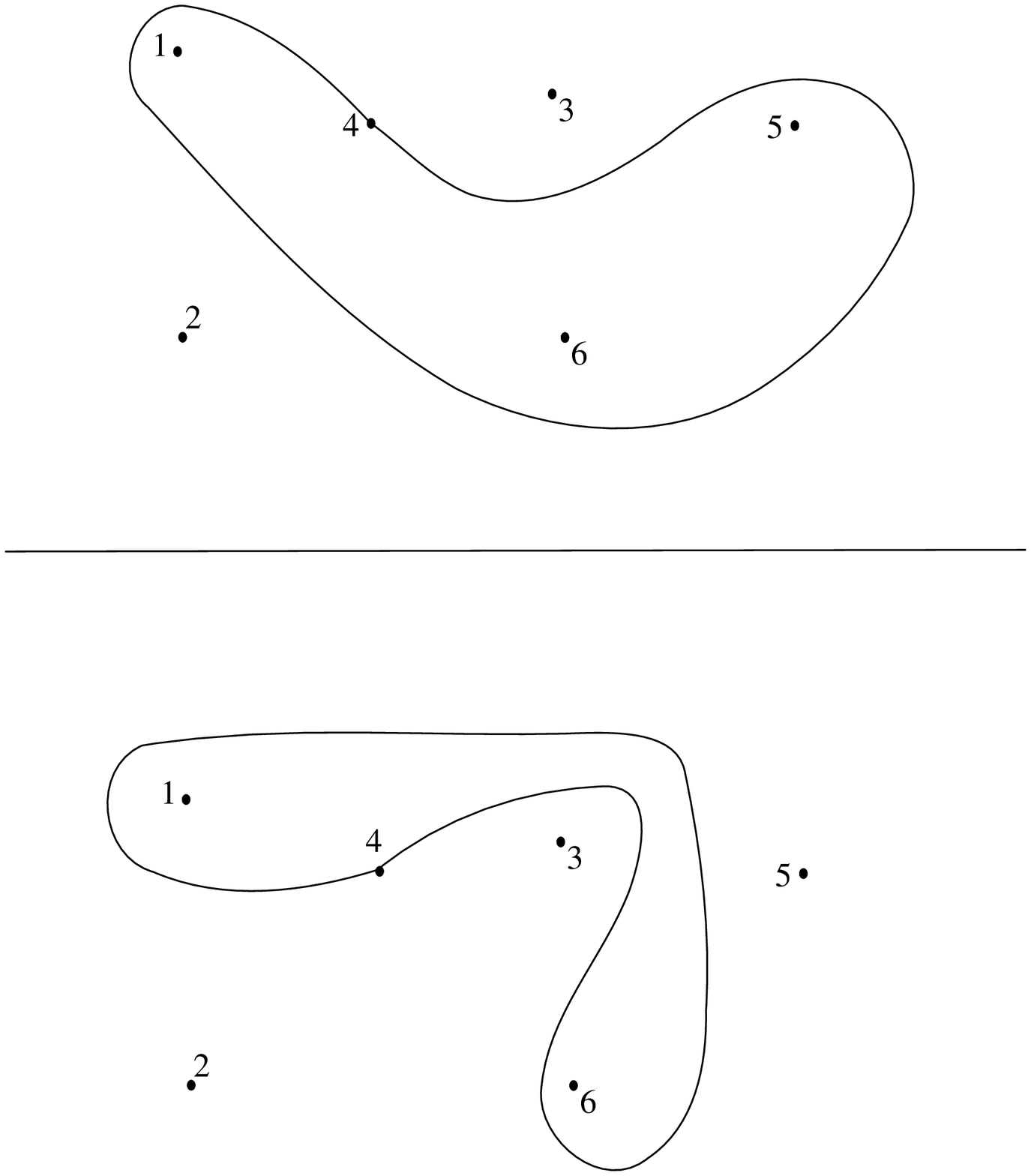}
\caption{All the stack-to-stack-strings with $\Lambda=-\mathbbm{1}$ that exist for $E_8$ but not for $E_6$ and $E_7$.}\label{fig:stackstackE8}
\end{figure}

We conclude with some brief remarks about $E_8$. The adjoint decomposes into irreps of $SU(5)\times (U(1))^4$ as
\begin{equation}
\mathbf{248}\rightarrow\mathbf{24}+4\cdot\mathbf{1}+5\cdot\left(\mathbf{10}+\bar{\mathbf{10}}\right)+
10\cdot\left(\mathbf{5}+\bar{\mathbf{5}}\right)+20\cdot\mathbf{1}\,,
\end{equation}
where the four singlets $4\cdot\mathbf{1}$ correspond to Cartan generators and the 20 singlets, $20\cdot\mathbf{1}$, can be divided into two set of opposite $U(1)$ charges, so that there are at least 10 single-to-single-brane-strings needed. A branch cut representation for a manifest $SU(5)$ inside the $[-(T^{-1}S)^2]$ loop is given in Fig.~\ref{fig:branchcutsEgroups}. It can be checked that now there are, besides the three homotopically inequivalent stack-to-stack-strings with $\Lambda=-\mathbbm{1}$ that exist in the $E_7$ case, two more such strings, see Fig.~\ref{fig:stackstackE8}. The number of stack-to-single-brane- and single-to-single-brane-strings that exist for $E_8$ but not for $E_7$ is rather large. Conditions I and II are trivially met because the number of stack-to-single-brane- and single-to-single-brane-strings in the $E_7$ case with $SU(5)$ manifest which is included in the $E_8$ case is already sufficiently high.

\section{Discussion}

Let us recapitulate the situation for the $E$-type symmetry groups. Due to the fact that we cannot identify the $U(1)$ charges of the open strings we cannot relate BPS strings to specific gauge group generators. The number of BPS strings with at least one endpoint on a single brane is generically larger than the number of gauge group generators. The number and type of BPS strings strongly depends on which symmetry group is made manifest and via which embedding this is done, i.e.~which branch cut representation is used.

We used homotopy inequivalence of BPS strings (except for the $[T^n]$ conjugacy classes) as a necessary condition to distinguish between different strings. A stronger condition would be to say that homotopically distinct BPS strings are only to be considered inequivalent when they have different $U(1)$ charges. Certainly, from the point of view of gauge group generators that should be sufficient. The incorporation of $U(1)$ charges into the analysis, however, remains an open problem.

In this work we have shown that it is conceivable that an open string interpretation of the $A$-$D$-$E$-type symmetry groups exists provided a sufficient number of branes is non-coinciding. We stress that we cannot follow step by step what happens  when an $A$-$D$-$E$-gauge group is actually formed with the exception of the $A$-groups corresponding to the $[T^n]$ conjugacy classes. This is either because the open string description breaks down once too much symmetry is made manifest or, more generically, because at some point one has to put 7-branes on top of locked 7-branes and such a process does not correspond to a continuous change of the branch cuts, so that it is not clear how to do this graphically. We therefore cannot trace the fate of the BPS strings all the way down to the formation of the singularity. It would be interesting to see how, with our global branch cut structure and with so much symmetry made manifest that ordinary open BPS strings have seized to exist, the open string description of the BPS states gets replaced by a description in terms of multi-pronged strings
\cite{Gaberdiel:1997ud,Gaberdiel:1998mv}.

Finally, in this work we have been using the Kodaira classification, see Table \ref{Tatealgorithm}, and reasoned our way towards the known singularity structures.  It would be rather satisfying to {\it{derive}} the Kodaira classification from the open string perspective presented here.

 \section*{Acknowledgments}
The authors wish to thank Radu Tatar and, in particular, Matthias Gaberdiel for useful discussions. We are also grateful to Teake Nutma for help with making the pictures. This work was supported in part by the Swiss National Science Foundation and the ``Innovations- und Kooperationsprojekt C-13'' of the Schweizerische Universit\"atskonferenz SUK/CRUS. J. H. wishes to thank the University of Groningen for its hospitality.

\end{document}